\documentclass[12pt]{article}
\pdfoutput=1
\usepackage{graphicx,psfrag,epsf,color}
\usepackage{amsmath,amssymb,amsfonts}
\usepackage{array}
\usepackage{cite}
\usepackage{rotating}
\usepackage{slashed, cancel}
\usepackage{xcolor}
\usepackage[caption=false]{subfig}
\usepackage{graphicx}
\usepackage{mathrsfs}  
\usepackage{tabularx}
\newcolumntype{C}{>{\centering\arraybackslash}X}

\newcounter{MBQ}

\newcommand{\eps}{\epsilon}

\newcommand{\wt}{\widetilde}

\def\be{\begin{equation}}
\def\ee{\end{equation}}
\def\beq{\begin{eqnarray}}
\def\eeq{\end{eqnarray}}
\newcommand{\bea}{\begin{eqnarray}}
\newcommand{\eea}{\end{eqnarray}}
\newcommand{\beas}{\begin{eqnarray*}}
\newcommand{\eeas}{\end{eqnarray*}}

\newcommand{\qboth}{\otimes}
\newcommand{\qtwo}{Q_2}
\newcommand{\alem}{\alpha_{\rm em}}

\newcommand{\GeV}{\,{\rm GeV}}

\newcommand{\bra}[1]{\big\langle{#1}\big\vert}
\newcommand{\ket}[1]{\big\vert{#1}\big\rangle}

\begin{document}
\allowdisplaybreaks

\begin{titlepage}

\begin{flushright}
{\small
TUM-HEP-1277/20\\
MPP-2020-160\\
August 23, 2020\\
}
\end{flushright}

\vskip1cm
\begin{center}
{\Large \bf\boldmath QED factorization of non-leptonic $B$ decays}
\end{center}

\vspace{0.5cm}
\begin{center}
{\sc Martin~Beneke,$^a$ Philipp B\"oer,$^a$ 
Jan-Niklas Toelstede,$^{a,b}$ K. Keri Vos$^{a}$} \\[6mm]
{\it $^a$Physik Department T31,\\
James-Franck-Stra\ss{}e~1, 
Technische Universit\"at M\"unchen,\\
D--85748 Garching, Germany\\[0.3cm]

$^b$Max-Planck-Institute for Physics, \\
F\"ohringer Ring 6, D-80805 Munich, Germany}
\end{center}

\vspace{0.6cm}
\begin{abstract}
\vskip0.2cm\noindent
We show that the QCD factorization approach for $B$-meson  
decays to charmless hadronic two-body final states can be extended 
to include electromagnetic corrections. The presence of 
electrically charged final-state particles complicates 
the framework. Nevertheless, the factorization formula takes 
the same form as in QCD alone, with appropriate 
generalizations of the definitions of light-cone distribution 
amplitudes and form factors to include QED effects. 
More precisely, we factorize QED effects above the strong 
interaction scale $\Lambda_{\rm QCD}$ for the non-radiative 
matrix elements
$\langle M_1 M_2|Q_i|\bar{B}\rangle$ of the current-current 
operators from the effective weak interactions. The 
rates of the branching fractions for the infrared-finite 
observables $\bar B\to M_1 M_2(\gamma)$ with photons of 
maximal energy $\Delta E\ll\Lambda_{\rm QCD}$ is then obtained by 
multiplying with the soft-photon exponentiation factors. We 
provide first estimates for the various electromagnetic corrections, 
and in particular quantify their impact on the $\pi K$ ratios and 
sum rules that are often used as diagnostics of New Physics.
\end{abstract}
\end{titlepage}




\section{Introduction}
\label{sec:intro}

In this paper we generalize the QCD factorization 
formula~\cite{Beneke:1999br,Beneke:2000ry}
\begin{equation}
 \left\langle M_1 M_2 | Q_i |\bar{B} \right\rangle = 
F_{B \to M_1} \!\times T^{\rm I}_i * \phi_{M_2} 
+  T^{\rm II}_i* \phi_{M_1} * \phi_{M_2} * \phi_B
\label{eq:QCDF}
\end{equation}
for non-leptonic $B$  decays into two light mesons to 
include QED. The formula is valid in the heavy-quark limit 
and expresses the matrix elements of operators $Q_i$ from 
the effective weak interactions below the electroweak 
scale in terms of $B\to M_1$ transition form factors, 
light-cone distribution amplitudes (LCDAs) of the $B$ meson and 
final mesons $M_{1,2}$, and their convolution 
with short-distance kernels. The latter can be computed 
in an expansion in the strong coupling $\alpha_s$. The 
first term in the formula is usually referred to as the 
``form-factor term'', the second one as the 
``hard spectator-scattering term''.

The QCD corrections to the short-distance kernels 
$T^{\rm I,II}_i$ are already known to 
$\mathcal{O}(\alpha_s^2)$ (NNLO)
\cite{Beneke:2005vv,Beneke:2006mk,Bell:2007tv,Bell:2009nk,Beneke:2009ek,Kim:2011jm,Bell:2015koa,Bell:2020qus}. Together with 
the expectation of high-precision measurements from LHCb and 
from the BELLE~II experiment at KEK, this motivates the consideration 
of QED effects despite the smallness of the electromagnetic 
coupling $\alpha_{\rm em}$. We shall present first estimates 
for a number of observables in this work. However, its main 
purpose is to investigate whether and how QED can be 
included in a factorization formula for non-leptonic $B$ decays. 
Quite generally, and perhaps contrary to intuition, the 
factorization of QED effects is more complicated than 
that of QCD, because the mesons are always colour-neutral, 
but can be electrically charged. Recent work on 
electromagnetic corrections to $B_s\to \mu^+\mu^-$ 
has shown \cite{Beneke:2017vpq,Beneke:2019slt} that they can 
manifest qualitatively new effects such as power-enhancement 
in the heavy-quark limit relative to the leading pure-QCD 
amplitude. While no such power-enhancement 
appears in the non-leptonic amplitudes discussed in this 
work, QED again leads to a number of effects not present 
in QCD alone, all related to the non-decoupling of soft 
photons from the electrically charged initial and final states. 
Although QED is weak, the interaction of photons with soft quarks 
is non-perturbative, which leads to a much more complicated 
structure of the hadronic matrix elements, required to 
account for QED corrections. The main results of this paper 
demonstrate that factorization for non-leptonic $B$ decays 
can be extended to include QED, provide operator definitions 
for the hadronic matrix elements, and give the short-distance 
QED kernels at $\mathcal{O}(\alpha_{\rm em})$. 

To put our discussion into a more general perspective, 
let us emphasize that the branching fraction for the 
decay $B\to M_1 M_2$ is not infrared-finite once QED 
corrections are included. Likewise the matrix elements 
$\langle M_1 M_2 | Q_i |\bar{B}\rangle$ are infrared 
divergent. The observable of interest is the 
branching fraction $B\to M_1 M_2 (\gamma)$, where 
$\gamma$ represents any number of soft photons with total 
energy less than $\Delta E$ in the $B$-meson rest frame, 
and we assume that $\Delta E\ll \Lambda_{\rm QCD}$, 
the scale of the strong interaction. QED effects above 
this ``ultrasoft'' scale $\Delta E$ are therefore purely 
virtual. What we compute for the first time in this paper are the QED 
corrections to the so-called non-radiative amplitude, 
which corresponds to the purely virtual contribution to 
the non-radiative process $B\to M_1 M_2$ with virtual 
corrections below the scale  of a few times $\Delta E$ 
removed.

A standard treatment of QED effects takes the pure-QCD 
amplitude and dresses it with Bloch-Nordsieck factors that 
exponentiate the large collinear and soft logarithms 
$\ln \frac{m_B}{m_{M}}$ and $\ln\frac{m_B}{\Delta E}$, 
respectively. This procedure is incomplete 
in several respects. The choice of the $B$-meson mass $m_B$
in the logarithm implies that the mesons are assumed to be 
point-like to distances of order $1/m_B$ instead of the 
true size of hadrons, $1/\Lambda_{\rm QCD}$. It also neglects 
electromagnetic effects above the scale $m_B$. While the latter 
can be taken into account in a conceptually straightforward 
way by including electromagnetic effects into the matching 
and evolution of the Wilson coefficients of the effective weak 
interaction operators $Q_i$, below the 
scale $m_B$ the situation becomes more complicated. As discussed 
in \cite{Beneke:2019slt}, between $m_B$ and a scale a few 
times $\Lambda_{\rm QCD}$, QED effects can be computed 
in the QED extension of the soft-collinear effective theory 
(SCET) framework, more precisely by the two-step 
matching to SCET$_{\rm I}$ and  SCET$_{\rm II}$. At the 
scale $\Lambda_{\rm QCD}$, SCET$_{\rm II}$ is strongly coupled 
but soft photons can still resolve the structure of the mesons.
Only at scales a few times $\Delta E \ll \Lambda_{\rm QCD}$, 
perturbative computations are again possible, since the 
mesons can now be treated as point-like particles in a 
multipole expansion, in which the leading interaction term is fixed 
by gauge invariance. In the present paper, we accomplish the 
systematic 
factorization and calculation of electromagnetic effects within 
SCET and therefore extend the rigorous computation of 
QED effects from $m_B$ down to scales of a few times $\Lambda_{\rm QCD}$. 
There remains a gap in our ability to compute QED effects 
related to the intrinsically non-perturbative effects at the 
scale $\Lambda_{\rm QCD}$, which prevent a perturbative 
matching of SCET$_{\rm II}$ to the effective theory of point-like 
hadrons.

The outline of the paper is as follows. 
In Section~\ref{sec:facformulae} we introduce some basic definitions 
and then immediately state the factorization formulas that include 
QED effects for the so-called current-current operators $Q_{1,2}$ 
in the effective weak interaction Lagrangian. The factorization 
formula takes the same form as in QCD alone. In the SCET 
formalism the short-distance information is contained in 
the hard-scattering kernels of SCET$_{\rm I}$ operators and 
the hard-collinear ``jet'' function from matching the 
spectator-scattering term to SCET$_{\rm II}$. We compute 
them at $\mathcal{O}(\alem)$ in QED in Sections~\ref{sec:SCET1} 
and~\ref{sec:spectatorscat}, respectively. However, compared 
to QCD, the non-perturbative objects in (\ref{eq:QCDF})---the 
decay constants, LCDAs and form factors---must be generalized 
to include QED effects. Their definition and renormalization 
is discussed in Section~\ref{sec:SCETIfact} and further in 
Section~\ref{sec:spectatorscat}, but more details on their 
renormalization group equations are left to \cite{BLCDApaper}. 
Section~\ref{sec:usoft} presents a treatment of the ultrasoft 
effects mentioned above at the leading logarithmic accuracy. 
We end with first estimates of QED effects in the colour-allowed 
and colour-suppressed tree amplitudes for $\pi K$ two-body final 
states in Section~\ref{sec:pheno}, and evaluate ratios 
of branching fractions that are often employed as diagnostics 
of New Physics. 
An Appendix rederives the spectator-scattering kernels in 
the ``old-fashioned'' projection formalism to clarify some 
subtleties in the interpretation of endpoint-singular 
convolutions.



\section{Factorization formulas}
\label{sec:facformulae}

In this work we consider the decay of a $B_q$ (with $q=u,d,s$) meson into two light pseudo-scalar mesons $M_1$ and $M_2$ mediated by the current-current operators for $b\to u$ transitions, 
given by the weak Hamiltonian
\begin{equation}             
\mathcal{H}_{\rm eff} = \frac{G_F}{\sqrt{2}} \,V_{uD}^* V_{ub}
           \left( C_1 Q_1 + C_2 Q_2\right) + \mathrm{h.c.}
\end{equation}
with the CMM operator basis~\cite{Chetyrkin:1997gb}  
\begin{align}
        Q_1 &= [\bar u \gamma^\mu T^a (1 -\gamma_5) b]
                [\bar D \gamma_\mu T^a (1 - \gamma_5) u] ,
\nonumber\\
        Q_2 &= [\bar u \gamma^\mu (1 -\gamma_5) b]
                 [\bar D \gamma_\mu (1 - \gamma_5) u],
\label{eq:weakham}
\end{align}
and $D=d$ or $s$. $T^a$ denotes the SU(3) colour generator.

\begin{figure}[t]
\centerline{\includegraphics[scale=1.1]{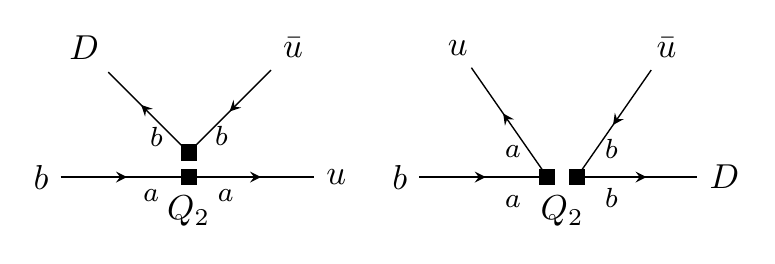}}
\caption{\label{fig:rwinsert} ``Right'' and ``wrong''
insertions of the operator $Q_2$, respectively.}
\end{figure}

We have to consider two possible flavour flows depicted in 
Fig.~\ref{fig:rwinsert} for $Q_2$ (see e.g. \cite{Beneke:2005vv}). 
First, the ``right'' insertion, where the ``emitted meson'' $M_2$ 
carries flavour $(D\bar{u})$ and is formed from the 
$[\bar{D} u]$ quark bilinear in $Q_{1,2}$ with spinor indices 
contracted in the bracket. This contributes to the colour-allowed 
tree-amplitude $\alpha_1(M_1 M_2)$.\footnote{Notation as in 
\cite{Beneke:2003zv}.} Second, the ``wrong'' insertion, 
which contributes to the colour-suppressed tree-amplitude 
$\alpha_2(M_1 M_2)$, in which case $M_2$ is made up of a 
$(u\bar u)$ pair from two different bilinears in $Q_{1,2}$. A Fierz 
transformation would be required in order to factorize the 
spinor index contractions into a $B\to M_1$ transition and 
a vacuum~$\to M_2$ transition. Contributions to penguin amplitudes 
from contractions of the $u$ and $\bar{u}$ field in the same 
fermion loop are not considered in this paper.

Our main result is that the 
QCD factorization formula can be extended to include 
QED corrections, and takes the same form as in pure QCD:
\begin{eqnarray}
\label{eq:QEDF}
\left\langle M_1 M_2 | Q_i |\bar B\right\rangle &=& i m_B^2\,
\bigg\{
\mathcal{F}^{BM_1}_{Q_2}(0) \, \int_0^1 du \, 
T^{{\rm I}}_{i,Q_{2}}(u) \mathscr{F}_{M_2} \Phi_{M_2}(u) 
\nonumber \\
&&\hspace*{-1.5cm}+ 
\,\int_{-\infty}^\infty d \omega \int_0^1 du \; dv \,
T^{{\rm II}}_{i, \qboth}(u,v,\omega) 
\mathscr{F}_{M_1} \Phi_{M_1}(v) \mathscr{F}_{M_2}\Phi_{M_2}(u) 
\mathscr{F}_{B, \qboth} \Phi_{B,\qboth} (\omega)
\,\bigg\} \,.
\end{eqnarray}
However, the short-distance kernels now depend on the 
electric charge $Q_2$ of $M_2$ or the charges of both mesons. 
In this case we use the symbol $\qboth =(Q_1,Q_2)$. In addition, 
all non-perturbative objects, the heavy and light meson's 
decay constants $\mathscr{F}$ and LCDAs 
$\Phi$, and form factors $\mathcal{F}^{BM_1}(0)$ at $q^2=0$,  
are generalized to include virtual long- and short-distance 
photon exchanges. In particular, the $B$-meson decay constants, 
LCDAs, and form factors become process-dependent. 

The QCD$\times$QED factorization formula thus describes the 
four different cases $\qboth = (0,0), (-,0), (0,-), (+,-)$. In the 
first two cases, where the meson $M_2$ emitted from the 
$B\to M_1$ transition is electrically neutral, only 
the ``wrong'' insertion of the operators $Q_{1,2}$ 
contributes. Since $M_2$ is colour- and charge-neutral, soft  
gluons and photons decouple completely from $M_2$ 
in the heavy-mass limit $m_b\to \infty$, and the 
situation closely resembles that of pure QCD. Moreover, the 
SCET$_{\rm I}$ $B\to M_1$ form factor can be related to and 
substituted by the full QCD$\times$QED $B \to M_1$ transition form 
factor as is usually done in pure QCD. The full QCD$\times$QED 
$B \to M_1$ form factor will be slightly different for a charged 
and a neutral meson transition due to QED effects. 

When the emitted meson $M_2$ is charged, corresponding to 
$\qboth =(0,-), (+,-)$ only the ``right'' operator insertion 
contributes, but the situation is more involved. Soft photon 
exchanges between $M_2$ and the $B\to M_1$ transition do not 
cancel and require introducing a process-dependent 
$B \to M_1$ transition ``form factor'' 
$\mathcal{F}^{BM_1}_{Q_2}(0)$ that knows about 
the electric charge and direction of flight of $M_2$. This 
generalized SCET$_{\rm I}$ $B\to M_1$ form factor will contain 
soft spectator-scattering contributions, 
which would otherwise result in endpoint-singular 
convolution integrals. As in the case of neutral $M_2$, 
the SCET$_{\rm I}$ form factor could be replaced by a 
QCD$\times$QED transition form factor. The relevant amplitude 
is the non-radiative semi-leptonic 
$\bar B \to M_1 \ell^- \bar{\nu}_\ell$ amplitude in the kinematic limit 
where the neutrino becomes soft, $q^2=0$ and $E_\ell = m_B/2$. 
We then replace\footnote{A precise formulation of this 
schematic replacement is given in (\ref{eq:semilepFF}).} 
\begin{align}
\mathcal{F}^{BM_1}_-(q^2=0) &\quad\to\quad \frac{1}{C_{\rm sl} 
Z_{\ell}}\times 
\mathcal{A}_{\bar B \to M_1 \ell^- \bar{\nu}_\ell}^{\text{non-rad}}(q^2=0, E_{\ell} = m_B/2) \,,
\end{align}
together with an appropriate redefinition of the 
hard scattering kernel $T^{\rm I}_i$, 
which follows from the QED factorization 
formula for the semi-leptonic transition, analogous to 
(\ref{eq:QEDF}). Here 
$Z_\ell$ is a lepton-vacuum matrix element of a local 
SCET operator \cite{Beneke:2019slt}, which appears as a remnant of 
collinear factorization after introducing the semi-leptonic 
amplitude (more details in Section~\ref{sec:softff}).

We derive the factorization formulas within the framework of  SCET 
\cite{Bauer:2000yr,Bauer:2001yt,Beneke:2002ph,Beneke:2002ni}
in the following sections. This can be done in a two-step matching 
procedure QCD$\times$QED $\to$ SCET$_{\rm I}$ $\to$ SCET$_{\rm II}$
(see \cite{Beneke:2015wfa} for a review of this approach 
for QCD factorization of non-leptonic decays). 
Along with this we give the operator definitions of all 
non-perturbative objects in QCD$\times$QED.
Further, we compute the $\mathcal{O}(\alpha_{\rm em})$ 
contributions to the scattering kernels $T_i^{\rm I,II}$.



\section{Matching onto SCET$_{\rm I}$, renormalization and hard-scattering kernels}
\label{sec:SCET1}

In the first matching step the current-current operators $Q_{i}$ 
are matched onto operators in ${\rm SCET}_{\rm I}$ by integrating 
out hard fluctuations at the scale $m_b$.
As the effective theory description is akin to the pure QCD case we mainly follow the conventions of \cite{Beneke:2005vv}, 
where the meson $M_1$ moves in the direction of the light-like reference vector $n_-^\mu$ and $M_2$ moves into the opposite direction $n_+^\mu$, with $n_+^2 = n_-^2 = 0$ and $n_+ n_- = 2$.
It is convenient to work in the $B$ rest frame in which the four-velocity of the $B$ meson is $v^\mu = \frac12 (n_+^\mu + n_-^\mu) = (1,0,0,0)$. 


\subsection{\boldmath ${\rm SCET}_{\rm I}$ operators}
\label{sec:SCETIops}

In pure QCD, the ${\rm SCET}_{\rm I}$ operators consist of an A0- and B1-type heavy-to-light current for the $B \to M_1$ transition~\cite{Beneke:2002ph, Beneke:2003pa} multiplied with the unique anti-collinear structure $[\bar{\chi}_{\bar C}(t n_-) \frac{\slashed{n}_-}{2} (1-\gamma_5) \chi_{\bar C}(0)]$ related to the leading-twist LCDA $\phi_{M_2}$ of the emitted meson. In QCD$\times$QED 
however, the flavours $u$ and $D$ are distinguishable due to their different electromagnetic coupling.
We thus introduce two copies of the effective operators depending on the charges of the final state quarks. Generalizing from the 
pure QCD case \cite{Beneke:2005vv},
the matching equation then takes the form
\begin{equation}
\label{eq:matchI}
 Q_i(0) = \int d\hat{t} \, \tilde{H}_{i,\qtwo}^{\rm{I}}(\hat{t}) {\cal O}^{\rm I}_{\qtwo}(t) + 
 \int d\hat{t} \, d\hat{s} \, \left[\tilde{H}^{{\rm II}\gamma}_{i,\qtwo}(\hat{t},\hat{s}) {\cal O}^{{\rm II}\gamma}_{\qtwo}(t,s) \, +\tilde{H}^{{\rm II}g}_{i,\qtwo}(\hat{t},\hat{s}) {\cal O}^{{\rm II}g}_{\qtwo}(t,s) \right]\,,
\end{equation}
with $\hat{t} = n_- q \, t = m_B t$, $\hat{s} = n_+ p' s = m_B s$, 
and $p' (q)$  the momentum of the $M_1(M_2)$ meson. The charge-dependent SCET$_{\rm I}$ operators are
\begin{align}
\label{eq:opdefscetI}
 {\cal O}^{\rm I}_{0}(t) &= [\bar{\chi}^{(u)}_{\bar C}(t n_-) \frac{\slashed{n}_-}{2} (1-\gamma_5) \chi^{(u)}_{\bar C}(0)] \, \bar{\chi}^{(D)}_C(0) \slashed{n}_+ (1-\gamma_5) \, h_v(0) \,, \nonumber \\
  {\cal O}^{\rm I}_{-}(t) &= [\bar{\chi}^{(D)}_{\bar C}(t n_-) \frac{\slashed{n}_-}{2} (1-\gamma_5) \chi^{(u)}_{\bar C}(0)] \, \bar{\chi}^{(u)}_C(0) \slashed{n}_+ (1-\gamma_5) \, h_v(0) \,, \nonumber \\
 {\cal O}^{{\rm II} \gamma}_{0}(t,s) &= \frac{1}{m_b}[\bar{\chi}^{(u)}_{\bar C}(t n_-) \frac{\slashed{n}_-}{2} (1-\gamma_5) \chi^{(u)}_{\bar C}(0)] \, \bar{\chi}^{(D)}_C(0) \, \frac{\slashed{n}_+}{2} \, \slashed{\cal A}_{C,\perp}(s n_+) (1+\gamma_5) \, h_v(0) \,, \nonumber \\
 {\cal O}^{{\rm II} \gamma}_{-}(t,s) &= \frac{1}{m_b}[\bar{\chi}^{(D)}_{\bar C}(t n_-) \frac{\slashed{n}_-}{2} (1-\gamma_5) \chi^{(u)}_{\bar C}(0)] \, \bar{\chi}^{(u)}_C(0) \, \frac{\slashed{n}_+}{2} \, \slashed{\cal A}_{C,\perp}(s n_+) (1+\gamma_5) \, h_v(0) \,, 
\end{align}
and ${\cal O}^{{\rm II}g}_{\qtwo}$ can be obtained by replacing $\cal{A} \to \cal{G}$. 
Here $\chi$ ($\mathcal{A}^\mu$, $\mathcal{G}^\mu$) are the 
collinear gauge-invariant building blocks in SCET for the 
collinear quark (photon, gluon) fields. Capital ``$C$'' denotes 
SCET$_{\rm I}$ collinear fields, which can have hard-collinear or 
collinear virtuality, while ``$c$'' refers exclusively to collinear 
virtualities (similarly, for the anti-collinear fields).
Gauge-invariance is achieved by dressing fields 
with the $SU(3)_c \times U(1)_{\rm em}$ collinear Wilson lines
\begin{equation}
\chi^{(q)}_{C} = [W_C^{(q)}]^\dagger \xi^{(q)}_{C} \,, \qquad \qquad
\chi^{(q)}_{\bar{C}} = [W_{\bar C}^{(q)}]^\dagger \xi^{(q)}_{\bar{C}} \,,
\end{equation}
where
\begin{align}
 W_{C}^{(q)} &= \exp \left\{ + i Q_q e \int_{-\infty}^0 d s \, n_+ A_{C}(x + s n_+) \right\}  \, 
 {\mathbf P} \exp \left\{ i g_s \int_{-\infty}^0 d s' \, n_+ G_{C}(x + s' n_+) \right\} \,.
\end{align}
$Q_q$ denotes the electric quark charge in units of 
$e = \sqrt{4 \pi \alem}$.
The SCET building blocks for the (electrically neutral) photon and gluon fields are
\begin{equation}
\mathcal{A}_{C,\perp}^\mu = e \left[ A_{C,\perp}^\mu 
- \frac{i \partial_\perp^\mu n_+ A_C}{i n_+ \partial} \right] \,, 
\qquad \qquad \mathcal{G}_{C,\perp}^\mu = W_C^{(0)\dagger} 
\left[i D_{C,\perp}^{\mu,(0)} W^{(0)}_C \right] \,,
\end{equation}
where $W^{(0)}_C$ denotes the QCD-only part of the Wilson line, and 
similarly for the covariant derivative,  $iD_{C,\perp}^\mu = i \partial_\perp^\mu + e A_{C,\perp}^\mu + g_s G_{C,\perp}^\mu$. For the 
anti-collinear fields analogous definitions and conventions 
apply with the replacements $C\to \bar{C}$, $n_\pm\to n_\mp$.

At this point, the main difference in the anti-collinear sector 
of the $M_2$ meson with respect to pure QCD is that in the product 
$\bar{\chi}_{\bar{C}}^{(D)} \left[\;\ldots\right] 
\chi_{\bar{C}}^{(u)}$ the QCD Wilson lines combine to a 
finite-length Wilson line, but the QED Wilson lines do not for 
charged mesons due to the different quark electric charges.
We note that---as in pure QCD---the operators ${\cal O}^{\rm II}$ are suppressed by one power of $\Lambda_{\rm QCD}/m_b$ with respect to the ${\cal O}^{\rm I}$. However, as is well-known, the form-factor and hard spectator-scattering terms contribute to the decay amplitude at the same order in the heavy-quark expansion. Hence, both operators are relevant after integrating out the hard-collinear scale and matching onto SCET$_{\rm II}$. 

At leading power the $C$ and ${\bar C}$ fields can only interact with soft modes via the exchange of eikonal gluons or photons.
These interactions can be removed from the ${\rm SCET}_{\rm I}$ 
Lagrangian by redefining the collinear and anti-collinear fields with 
soft Wilson lines 
\begin{eqnarray}
\label{eq:softwilsonlines}
S^{(q)}_{n_\pm}(x) &=& \exp \left\{ -i Q_q e \int_0^{\infty} \!ds \,
 n_\pm A_{s}(x + s n_\pm) \right\}  \, 
{\mathbf P} \exp \left\{ -i g_s \int_0^{\infty} \!d s \, n_\pm 
G_s(x + s n_\pm) \right\} .
\nonumber\\[-0.2cm]
\\[-1.2cm]
\nonumber
\end{eqnarray}
If $\chi^{(q)}_{\bar C}$ creates an outgoing antiquark with electric charge $Q_q$, the redefinition reads 
\begin{align}
 \chi^{(q)}_{\bar C}(x) &\to S^{(q)}_{n_+}(x_+) \chi_{\bar C}^{(q)}(x) \,,
\end{align}
while $S^\dagger$ must be used for an outgoing quark.
As a consequence the anti-collinear meson $M_2$ decouples from the $B\to M_1$ transition already at the hard scale $m_b$.
As we only consider colour-singlet operators, the QCD part of the soft Wilson lines from the anti-collinear sector cancels.
However, the QED Wilson lines combine to a soft Wilson line $S_{n_+}^{\dagger (Q_{M_2})}(x_+)$ that carries the total electric charge of the emitted $M_2$ meson.


\subsection{Matching equation and renormalization}
\label{sec:matcheq}

\begin{figure}[t]
\begin{center}
\includegraphics[scale=1.00]{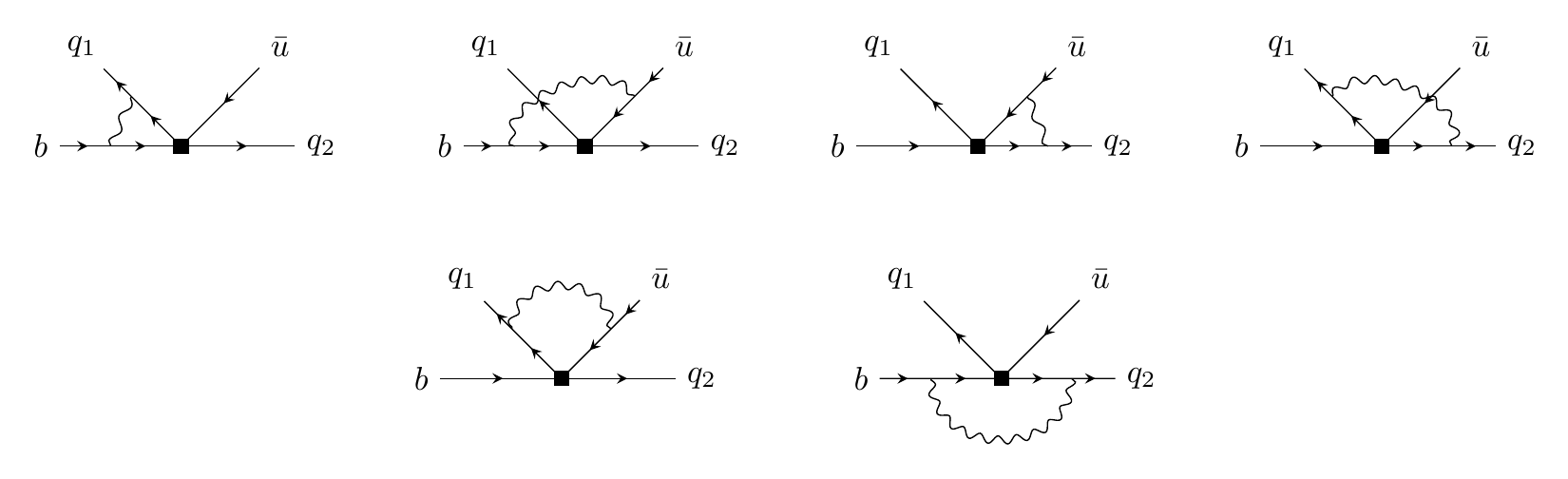}
\caption{$\mathcal{O}(\alem)$ vertex corrections to $H^{\rm{I}}$.}
\label{fig:diaI}
\end{center}
\end{figure}

\begin{figure}[t]
\begin{center}
\includegraphics[scale=0.95]{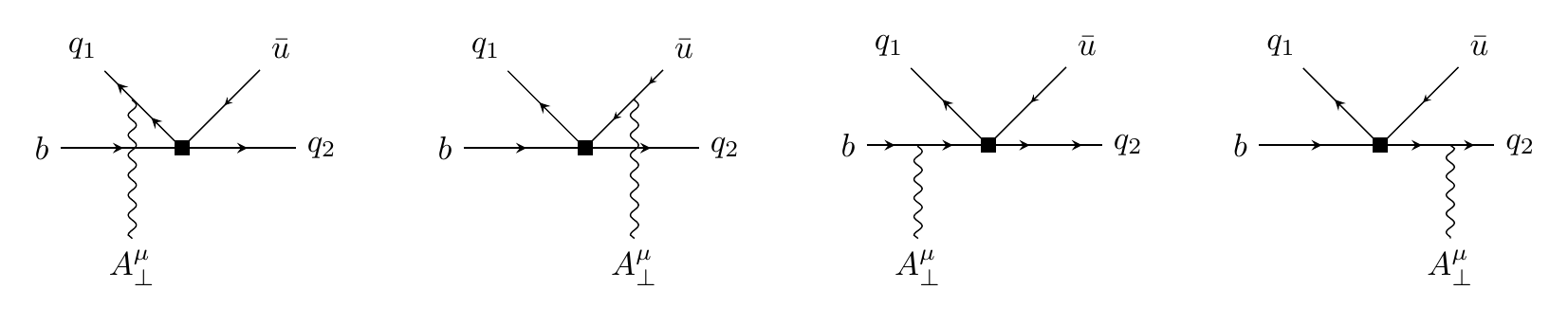}
\caption{$\mathcal{O}(\alem)$ spectator-scattering corrections to $H^{\rm{II}}$.}
\label{fig:diaII}
\end{center}
\end{figure}

We compute the matching coefficients 
$H_{i,\qtwo}^{\rm{I}}(u)$ and 
$H_{i,\qtwo}^{\rm{II}\gamma}(u,v)$ in (\ref{eq:matchI}) 
to $\mathcal{O}(\alem \alpha_s^0)$ by computing 
suitable quark-gluon matrix elements. More precisely, 
we compute the corresponding momentum-space 
coefficients.\footnote{As indicated by the omission of the 
tilde symbol. The relation reads $H^{\rm I}(u) = \int d\hat{t} \, e^{i u \hat{t}} \tilde{H}^{\rm I}(\hat{t}) $ and $H^{\rm II}(u,v) = \int d\hat{t}d \hat{s} \, e^{i (u \hat{t}+ (1-v) \hat{s})} \tilde{H}^{\rm II}(\hat{t}, \hat{s})$. } 
At this order, this implies one-loop QED matching of 
the $\mathcal{O}^{\rm I}$ operators, and 
tree-level matching of $\mathcal{O}^{{\rm II}\gamma}$. 
The diagrams are shown in Figs.~\ref{fig:diaI} and \ref{fig:diaII}. 
The matching coefficients 
$H_{i,\qtwo}^{\rm{II}g}$ start at 
$\mathcal{O}(\alpha_s)$ and correspond to the pure-QCD 
coefficients at this order.

In analogy to QCD, the ultraviolet (UV) renormalized matrix elements 
of the $Q_i$ can be written as
\bea
\left\langle{Q_i}\right\rangle &=& \bigg\{ A_{i}^{(0)} + \frac{\alem}{4\pi} 
\left[ A_{i}^{(1)} + Z_{\rm ext}^{(1)} \, A_{i}^{(0)} + 
Z_{ij}^{(1)} A_{j}^{(0)}\right]  + 
\mathcal{O}(\alem^2) \bigg\} \, \left\langle{{\cal O}}\right\rangle^{(0)} \ ,
\label{eq:QCDside}
\eea
where the superscript indicates the expansion coefficients 
in powers of $\alem(\mu)/(4\pi)$. 
Here $A_{i}^{(0)} (A_{i}^{(1)})$ are the bare tree-level (one-loop) 
on-shell matrix elements of operators $Q_i$. The factor 
\begin{align}
  Z_{\rm ext}^{(1)} = -\frac{1}{2} Q_d^2 \left(3\left[\frac{1}{\epsilon} + \ln\left(\frac{\mu^2}{m_B^2}\right)\right] +4 \right), 
\label{eq:OSZ} 
\end{align}
accounts for the one-loop on-shell renormalization of the $b$-quark  
field. The sum over $j$ includes not only the operators $Q_{1,2}$, 
but also the two evanescent operators (non-vanishing only 
in $d\neq4$ space-time dimensions) \cite{Beneke:2009ek}
\bea \label{eq:eva}
E_1^{(1)} &=&\bar u \gamma^\mu\gamma^\nu\gamma^\rho T^a (1-\gamma_5) b \;\, 
\bar D \gamma_\mu\gamma_\nu\gamma_\rho T^a (1-\gamma_5) u - 16 Q_1 \, ,
\nonumber\\
E_2^{(1)} &=&\bar u \gamma^\mu\gamma^\nu\gamma^\rho (1-\gamma_5) b \;\, 
\bar D \gamma_\mu\gamma_\nu\gamma_\rho (1-\gamma_5) u- 16 Q_2  
\eea
to close the operator basis under renormalization at the one-loop 
order.
The factor $Z_{ij}$ denotes the one-loop QED operator 
renormalization constants 
\begin{equation}\label{eq:Zij}
Z_{ij}^{(1)} = \frac{1}{\epsilon} \left(
\begin{array}{cccccc}
\;  6 Q_{u} Q_{d} \; & \; 0 \; &  \; \tfrac{1}{4}(Q_{u}^2 + Q_{d}^2 + 2 Q_{u} Q_{d})  \; & \;  0 \; \\
\; 0 \;  &     \;   6 Q_{u} Q_{d}   \;    &    \;      0    \;   & \;    \tfrac{1}{4}(Q_{u}^2 + Q_{d}^2 + 2 Q_{u} Q_{d})   \; \end{array}
\right), 
\end{equation}
where the column index $j$ refers to $(Q_1,Q_2,E_1^{(1)},E_2^{(1)})$. 
The evanescent operators contribute finite terms from  
$Z_{ij}^{(1)} A_{j}^{(0)}$ to $\left\langle{Q_i}\right\rangle$ 
through the $\mathcal{O}(\epsilon)$ terms of $A_{j}^{(0)}$. 
Eq.~(\ref{eq:QCDside}) applies to the matching onto 
both SCET$_{\rm I}$ operator types, $\mathcal{O}^{\rm I}$ 
and  $\mathcal{O}^{\rm II}$, but for the second the equation 
is trivial, since we need to consider only tree-level matching.

To obtain the UV finite hard-scattering kernels, 
we also need the renormalized matrix elements of the 
SCET operators. Since the on-shell matrix elements are scaleless 
and vanish, only the ultraviolet renormalization 
kernel $Y^{(1)}$ of the SCET${}_{\rm I}$ operator has to be included.
We then find for the hard-scattering kernels 
\bea \label{eq:masterform}
H_i^{(0)} &= &A^{(0)}_{i}\, , \nonumber \\ 
H_i^{(1)} &= &A^{(1)}_{i}+ Z_{ij}^{(1)} \, A^{(0)}_{j} + \left(Z_{ext}^{(1)}   - Y^{(1)}\right) A_{i}^{(0)}\, ,
\eea
where we have omitted the charge ($Q_2$) and SCET operator (I/II)
indices for simplicity. For the wrong insertion, the operators $Q_{i}$ match onto the Fierz-transformed operator $\tilde{\cal{O}}$ (see \cite{Beneke:2009ek}) which is equivalent to ${\cal{O}}$ in $d=4$ dimensions. In this case an additional term appears on the right-hand side of (\ref{eq:masterform}) from the requirement that the renormalized matrix element of the evanescent operator $\tilde{\cal{O}}-{\cal{O}}$ vanishes when 
infrared (IR) divergences are regulated with a non-dimensional regulator. At the one-loop order only the difference between the SCET renormalization kernels $\tilde{Y}^{(1)}-Y^{(1)}$ enters. However, this can be shown 
to be $\cal{O}(\epsilon)$, hence \eqref{eq:masterform} applies to both the right and wrong insertion. 

Since the anti-collinear fields are decoupled from the collinear and 
soft ones, we can write the SCET renormalization kernel as the sum of two pieces
\begin{equation}
\label{eq:Y1}
    Y^{(1)}(u, v) = Z_J^{(1)}\delta(u-v) + Z_{\bar{C}}^{(1)}(u, v) \ ,
\end{equation}
where $Z_{\bar{C}}$ is the anti-collinear kernel and $Z_J$ the SCET heavy-to-light current renormalization constant. These correspond, respectively, to the pole parts of anti-collinear loops and soft plus collinear loops.
In pure QCD, this expresses the factorization of the $M_2$ meson from the $B\to M_1$ transition, and the above SCET renormalization kernel indeed factorizes into two separately well-defined pieces. In QED, the situation is more involved, since soft photons connect $M_2$ and the $B\to M_1$ transition, when $M_2$ carries electric charge.

\subsubsection{Anti-collinear kernel}
\label{sec:QEDERBL}

\begin{figure}[t]
\centering
 \includegraphics[scale=1.05]{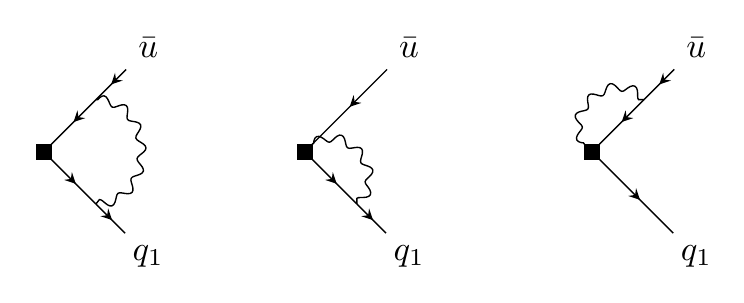}
 \caption[]{Diagrams at $\mathcal{O}(\alem)$ that contribute to the anomalous dimensions of the operator~\eqref{eq:QEDpiminusLCDA}.
 }
 \label{fig:QEDERBL}
\end{figure}

We first consider the QED renormalization of the anti-collinear 
operator
\begin{equation}
\label{eq:QEDpiminusLCDA}
 \bar{\chi}^{(q_1)}_{\bar{C}}(t n_-) \frac{\slashed{n}_-}{2}(1-\gamma_5) \chi^{(u)}_{\bar{C}} (0) \ .
\end{equation}
To this end we compute the diagrams in Fig.~\ref{fig:QEDERBL} by calculating its matrix element with external quark states with a small off-shellness ($k_{q_1}^2$ for the $q_1$-quark and $k_{\bar u}^2$ for the ${\bar u}$-quark) to ensure that all poles arise from UV divergences. Including  $\overline{\rm{MS}}$ external quark-field renormalization, we find 
\begin{equation}
 \langle \mathcal{O_{\rm bare}} \rangle^{\text{1-loop}}(u) = \frac{\alem(\mu)}{4\pi} \, \frac{2}{\epsilon} \, \int_0^1d v \, V(u,v) \, \langle \mathcal{O_{\rm bare}} \rangle^{\text{tree}}(v) + \mathcal{O}(\epsilon^0)
\end{equation}
with 
\begin{align}
\label{eq:UVpolescbar}
 V(u,v) &= \delta(u-v) \left( (Q_{q_1} - Q_u)^2 \left(\frac{1}{\epsilon} + \frac34 \right) + (Q_{q_1} - Q_u) \left( Q_{q_1} \ln\frac{\mu^2}{-k_{q_1}^2} - Q_u \ln \frac{\mu^2}{-k_{\bar u}^2} \right) \right) \nonumber \\ 
 &+ Q_u Q_{q_1} \left[ \left(1+\frac{1}{v-u}\right) \frac{u}{v}\, \theta(v-u) + \left(1+\frac{1}{u-v}\right) \frac{1-u}{1-v} \,\theta(u-v) \right]_+ \,.
\end{align}
The one-loop $Z$-factor is then given by
\begin{equation}\label{eq:ZBL}
Z^{(1)}_{\bar{C}}(u,v) =  
 \, -\frac{2}{\epsilon} \, V(u,v) \ .
 \end{equation}
The plus-distribution (in the variable $u$) is defined as
\begin{equation}
 \int_0^1 du \, \Big[ \dots \Big]_+ f(u) \equiv
 \int_0^1 du \, \Big[ \dots \Big] (f(u) - f(v))\,.
\end{equation}
For an electrically neutral meson ($q_1=u$) the first line in~\eqref{eq:UVpolescbar} vanishes and we recover the QCD ERBL evolution kernel~\cite{Lepage:1979zb, Lepage:1980fj,Efremov:1979qk} for the 
LCDA of pseudoscalar mesons upon replacing 
$ \alem Q_u^2 \to \alpha_s C_F$. However, for unequal quark 
charges, as applicable to an electrically charged meson, the 
$Z$-factor and corresponding anomalous dimension / kernel 
depend on the off-shellness of the quarks, that is, the IR 
regularization. We shall see next that this dependence is cancelled 
in the renormalization of the full SCET${}_{\rm I}$ operator, 
but take note that the above result implies that 
the anti-collinear operator (\ref{eq:QEDpiminusLCDA}) 
alone is ill-defined for unequal 
quark electric charges. We further note that the $1/\epsilon$ 
pole in $V(u,v)$ implies that $Z^{(1)}_{\bar{C}}(u,v)$ contains 
a double-pole, contrary to the corresponding QCD LCDA 
kernel, and hence the anomalous dimension has a cusp 
logarithm.


\subsubsection{Generalized heavy-to-light current}
\label{sec:QEDZJ}

The remaining soft and collinear fields of 
the $\mathcal{O}^{\rm I}$ operators 
define the generalized heavy-to-light current 
\begin{equation}
\label{eq:SCETIbm1op}
\bar{\chi}^{(q_2)}_{{C}}(0) \slashed{n}_+(1-\gamma_5) S_{n_+}^{\dagger(Q_{M_2})} h_v (0) \ .
\end{equation}
The soft Wilson line $S_{n_+}$ arises from the soft decoupling of 
the anti-collinear fields composing (\ref{eq:QEDpiminusLCDA}). We 
emphasize that no soft decoupling field redefinition has been 
performed in the collinear sector.\footnote{Soft Wilson lines 
without position argument are understood to refer to $x=0$.}

To calculate the renormalization factor $Z_J$ at the one-loop 
order, we regularize the IR divergences non-dimensionally by 
introducing an off-shellness $k_{q_2}^2$ for the light quark $q_2$.  
In addition, we have to modify the integer-charge soft Wilson line 
propagators as discussed in Appendix A in~\cite{Beneke:2019slt} 
to be consistent with the off-shell IR regulator used in the 
anti-collinear sector. For incoming photon momentum $k$, the 
soft Wilson-line propagator must be modified as
\begin{equation}\label{eq:propmod}
1/[n_+ k - i 0^+]  \to 1/[n_+ k -{\delta}_{\bar c} - i 0^+] \,,
\end{equation}
where
\begin{equation}
\label{eq:deltadef}
\delta_{\bar c}\equiv k_{q_1}^2/(n_-k_{q_1}) = 
k_{\bar{u}}^2/(n_-k_{\bar{u}})\,.
\end{equation}
The last equality imposes a relation between the a priori 
independent off-shellnesses $k_{q_1}^2$, $k_{\bar{u}}^2$ and 
momentum fractions of the quark and anti-quark of the 
anti-collinear meson $M_2$, which appear in (\ref{eq:UVpolescbar}). 
This relation is necessary to 
maintain the identity $S_{n_+}^{\dagger (d)} S_{n_+}^{(u)} = 
S_{n_+}^{\dagger (Q_{M_2})}$ for the regularized Wilson lines, which 
was used to obtain (\ref{eq:SCETIbm1op}) from the soft decoupling 
in the anti-collinear sector. 

\begin{figure}[t]
\centering
 \includegraphics[scale=1.10]{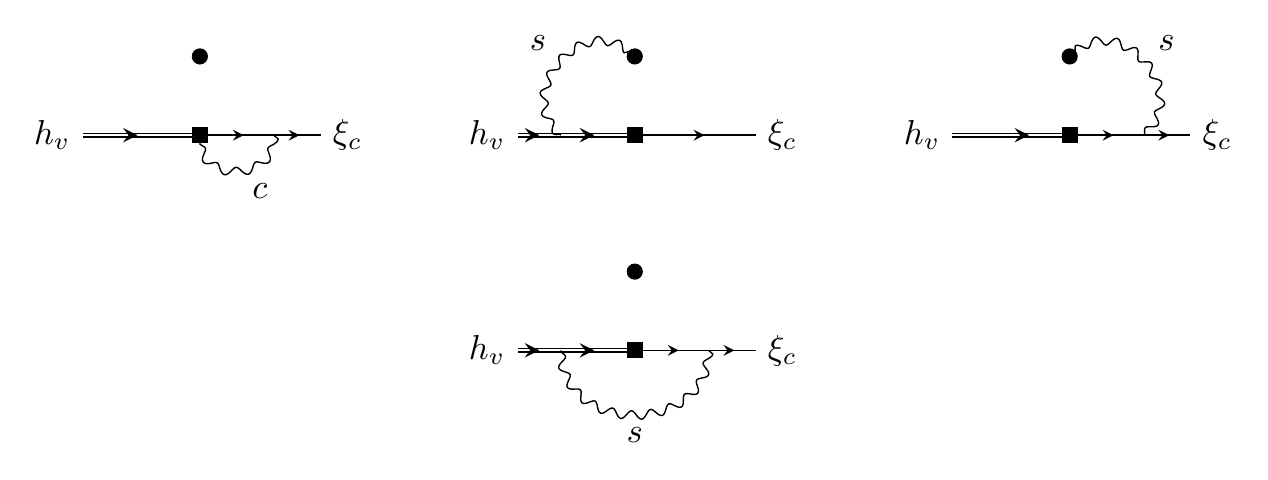}
\caption[]{Diagrams at $\mathcal{O}(\alem)$ that contribute to 
the renormalization of the operator~\eqref{eq:SCETIbm1op}. The 
black dot denotes the Wilson line operator 
$S_{n_+}^{\dagger(Q_{M_2})}$. Field-renormalization diagrams 
are not displayed.}
 \label{fig:ZJdias}
\end{figure}

The computation of the one-loop diagrams in Fig.~\ref{fig:ZJdias} 
gives the $\overline{\rm MS}$ renormalization factor  
\begin{eqnarray}
\label{eq:ZJ}
Z_{J}^{(1)}&=& - Q_d^2 \left\{\frac{1}{\epsilon^2} + 
\frac{1}{\epsilon}\left[L+\frac{5}{2}\right]  \right\} + 2Q_{M_2} Q_d\left\{\frac{1}{\epsilon^2} + 
\frac{1}{\epsilon}\left[L+\frac{3}{2} + i \pi\right]  \right\} \nonumber \\
&& - \,Q_{M_2}^2 \frac{1}{\epsilon}
\left[L+\frac{3}{2} + 2\ln\left(\frac{-\delta_{\bar c}}{\mu}\right) 
+ i \pi\right] ,
\end{eqnarray}
where we defined
\begin{equation}
    L \equiv \ln\left(\frac{\mu^2}{m_B^2}\right)\,.
\end{equation}
For neutral mesons $M_2$, that is for $Q_{M_2}=0$, this reduces to 
the QCD result (see e.g. \cite{Beneke:2009ek}) after replacing 
the charge factor $Q_d^2\to C_F$. 

\subsubsection{\boldmath SCET$_{\rm I}$ renormalization constants}
\label{sec:OIrenormalization}

Combining (\ref{eq:ZJ}) with $Z_{\bar{C}}^{(1)}$ 
from~\eqref{eq:ZBL} gives for the renormalization 
constant (\ref{eq:Y1}) of the SCET$_{\rm I}$ 
operators $\mathcal{O}^{\rm I}$
\begin{align}
Y^{(1)}(u,v) &  =
\delta(u-v)\,\Bigg(\! - Q_d^2 \left\{\frac{1}{\epsilon^2} + 
\frac{1}{\epsilon}\left[L+\frac{5}{2}\right]  \right\} 
+  2 Q_{M_2} Q_u\left\{ \frac{1}{\epsilon^2} 
+\frac{1}{\epsilon}\left[ L+i\pi+\frac{3}{2}  \right] \right\}  
\nonumber \\
& + \frac{2}{\epsilon}Q_{M_2}\Big[Q_d \ln{u}  
- Q_u \ln{(1-u)}\Big]\, \Bigg)
\\  
 & -\frac{2}{\epsilon} Q_u\left(Q_u+Q_{M_2}\right) \left[ \left(1+\frac{1}{v-u}\right) \frac{u}{v} \theta(v-u) + \left(1+\frac{1}{u-v}\right) \frac{1-u}{1-v} \theta(u-v) \right]_+\,.
\nonumber %
\label{eq:Y11}
\end{align}
As required, the full renormalization kernel does not depend on 
the IR regularization. However, as mentioned before, in QED  
the kernel does not factorize into an anti-collinear part 
and a soft plus collinear part, which could be employed to 
renormalize consistently the corresponding operators, since 
the separate anomalous dimensions would not be IR finite. This 
was already discussed previously \cite{Beneke:2019slt} for 
$B_q \to \mu^+\mu^-$ decays. Factorization can be restored by 
performing a rearrangement of soft-overlap terms, 
as will be discussed in Section~\ref{sec:SCETIfact}. 

The renormalization constant $Y^{(1)}$ of the SCET$_{\rm I}$ 
operators $\mathcal{O}^{\rm II}$ follow without further 
computation. The anti-collinear part is identical to  
(\ref{eq:QEDpiminusLCDA}) above, while the remaining soft and 
collinear fields are
\begin{equation}
\label{eq:SCETIOII}
\frac{1}{m_b}\bar{\chi}^{(q_2)}_C(0) \, \frac{\slashed{n}_+}{2} \, \slashed{\cal A}_{C,\perp}(s n_+) (1+\gamma_5) S_{n_+}^{\dagger(Q_{M_2})} h_v(0)\,.
\end{equation}
At the one-loop order, the renormalization constant of this 
operator coincides with 
(\ref{eq:ZJ}) for the simpler operator (\ref{eq:SCETIbm1op}), since the 
photon has no self-interactions and the renormalization 
of the $\bar{\chi}^{(q_2)}_C(0) \,\slashed{n}_+  
\Gamma S_{n_+}^{\dagger(Q_{M_2})} h_v(0)$ operator is independent 
of the Dirac matrix $\Gamma$. However, 
as we compute the matching coefficient of the  $\mathcal{O}^{\rm II}$ 
operators only at tree level, we will not need this result.


\subsection{\boldmath 
Hard-scattering kernels $H^{\rm {I}}_{i,\qtwo}$}
\label{sec:HI}

Matching $Q_{1,2}$ to the SCET operators 
$\mathcal{O}_{\qtwo}^{\rm I}(t)$ at tree-level gives 
the kernels $H_{i,\qtwo}^{\rm{I}(0)}$:
\bea 
H_{1,-}^{\rm{I}(0)}(u) &=& 0\,,\quad
H_{1,0}^{\rm{I}(0)}(u) = \frac{C_F}{N_c}\,,
\\
H_{2,-}^{\rm{I}(0)}(u) &=& 1\,,\quad
H_{2,0}^{\rm{I}(0)}(u) = \frac{1}{N_c}\,.
\eea
We recall that the right (wrong) insertion corresponds the the 
charge $Q_{M_2}=-1$ ($Q_{M_2}=0$) of $M_2$. 
The tree-level kernels coincide with those of pure QCD.
As in QCD the right insertion of the colour-octet operator $Q_1$ 
cannot match onto the colour-singlet SCET operator 
$\mathcal{O}_{\qtwo}^{\rm I}(t)$ at tree-level. This 
remains true in QED to all orders in $\alem$ (but 
leading order in $\alpha_s$), as does the relative 
colour factor between $H_{1,0}^{\rm{I}}(u)$ and 
$H_{2,0}^{\rm{I}}(u)$, that is
\begin{equation}
H_{1,-}^{\rm{I}}(u) = 0, 
\qquad H_{1,0}^{\rm{I}}(u) = C_F  H_{2,0}^{\rm{I}}(u)
\label{eq:pureQEDrelations}
\end{equation}
to all orders in pure QED. 

The hard-scattering kernels at the one-loop order 
can be extracted from the diagrams shown in 
Fig.~\ref{fig:diaI}. We compute the on-shell matrix elements 
$A_i^{(1)}$ and use the previously given renormalization factors  
to obtain  $H_{i,\qtwo}^{\rm{I}(1)}$ using~\eqref{eq:masterform}. 
In complete generality, we find for the right-insertion 
of the operator 
\begin{equation}
\label{eq:generalQ}
[\bar{q}_2 \gamma^\mu (1 -\gamma_5) b]
                 [\bar{q}_1 \gamma_\mu (1 - \gamma_5) u]
\end{equation}
the hard-scattering function
\begin{align}
    \label{eq:Hres}
   H^{\rm{I}(1)}_{2,-}(u) & = Q_{q_1} Q_{q_2} \left( L^2-4L_\nu+ L\left(4+2i\pi-2\ln u\right) + \ln^2{u} -2 i \pi \ln{u}  -  \frac{7\pi^2}{6}+1\right) \nonumber \\ %
    & -  Q_u Q_{q_2} \left( L^2 -L_\nu +L\left(4+2i\pi-2\ln\bar{u}\right)  - \ln{\bar{u}}\left(3+2i\pi - \ln\bar{u}\right)     - \frac{7\pi^2}{6} +3i\pi +6\right) \nonumber \\ %
    & +  Q_u Q_d \left( \frac{1}{2} L^2 -4L_\nu - 2 L\left(-1+\ln{\bar{u}}\right)+2\ln^2\bar{u} - \frac{2}{u} \ln{\bar{u}}   +2 {\rm Li}_2\left(u\right)  + \frac{\pi^2}{12} -3\right) \nonumber \\
    & -  Q_d Q_{q_1} \left( \frac{1}{2} L^2  -L_\nu + L(2-2\ln{u})+2 \ln^2{u}- 3\ln{u} + \frac{\ln{u}}{\bar{u}}  + 2 {\rm Li}_2(\bar{u}) + \frac{\pi^2}{12}+2  \right)  \nonumber \\
    &-3\left(Q_{q_1} + Q_u\right)\left(Q_{q_2} + Q_d\right)   \nonumber \\
    &- Q_{q_2} Q_d \left( \frac{1}{2} L^2 -L_\nu+ 2L + \frac{\pi^2}{12} +4\right)  - Q_d^2 \left(\frac{1}{2}L_\nu + L +2\right) 
-\frac{1}{2} Q_{q_2}^2 \left(L_\nu - L\right)\nonumber \\
    &-\frac{1}{2}\left(Q_{q_1}^2 + Q_u^2 - 2 Q_u Q_{q_1}\right)\left(L_\nu - L\right)\,,
   \end{align} 
where $Q_d$ represents the bottom-quark charge. We use 
the bar-notation $\bar u\equiv 1-u$ and introduced 
\begin{equation}
    L_\nu \equiv  \ln\left(\frac{\nu^2}{m_B^2}\right)\,,
\end{equation}
where $\nu$ refers to the scale of the Wilson coefficients, 
$C_i(\nu)$, which we distinguish from the scale $\mu$ in $L$. The 
explicit logarithms of $\nu$ cancel the electromagnetic scale 
dependence of the Wilson coefficients, whereas the $\mu$
dependence cancels with the scale dependence of the 
non-perturbative objects on the right-hand side of the 
factorization formula $(\ref{eq:QEDF})$ for the operator 
matrix elements such that the matrix elements 
are only $\nu$ dependent. 

The first four lines of (\ref{eq:Hres}) correspond to the first four 
diagrams in Fig.~\ref{fig:diaI} and together with the 
evanescent-operator contribution in the fifth line reproduce the QCD 
result \cite{Beneke:2009ek} for $T_1^{\rm I}(u)$ up to the colour 
factor $C_F/(2 N_c)$ when putting all quark electric charges equal 
to 1. The second-to-last line accounts for the last diagram in the 
second row of Fig.~\ref{fig:diaI} and $Z_{\rm ext}$, and---with the 
same replacement of charge factors---equals the quantity  
$C_{FF}/C_F$ in QCD as defined in \cite{Beneke:2009ek}.

The wrong insertion of the generalized four-quark operator 
(\ref{eq:generalQ}) can be obtained from 
the right insertion via
\begin{equation}
\label{eq:HresWI}
    H_{2,0}^{{\rm I}(1)}(u) = \frac{1}{N_c}H_{2,-}^{\rm I(1)}(u) - \frac{1}{N_c}\left(Q_d-Q_u\right)\left(Q_{q_2}-Q_{q_1}\right).
\end{equation}
For general quark charges the operator (\ref{eq:generalQ}) 
is not gauge-invariant, hence the coefficients of the 
different quark charge factors in (\ref{eq:Hres}) are 
gauge-dependent. The above result is given in Feynman gauge. 
The gauge dependence of course cancels when we specialize 
to the physical quark charge assignments as done next. We further 
note that above and in the remainder of this paper, we do not 
distinguish between the pole $b$-quark mass $m_b$ and the 
$B$-meson mass $m_B$, which hence appears in the argument of 
$L$. 

For the physical case when the meson $M_2$ is charged 
($q_1=d$ and $q_2=u$), we replace $Q_{q_1}=Q_d=-1/3$ and 
$Q_{q_2}=Q_{u}=2/3$ in (\ref{eq:Hres}) and obtain 
\begin{eqnarray}
H^{\rm{I}(1)}_{2,-}(u) & = &- \frac{13 L^2}{18}  +\frac{4}{3}L_\nu -L\left(\frac{41}{18} + \frac{4i\pi}{3}- \frac{4}{3}\ln(1-u) - \frac{2}{3}\ln u\right)  \nonumber \\
     &&- \frac{2 f(u) + 4f(1-u)}{9}- \frac{(2-u) \ln(u)}{3(1-u)}   + \frac{83\pi^2}{108}-\frac{4i\pi}{3}-\frac{19}{9}\,.
\label{eq:H2miphys}
\end{eqnarray}
Likewise for neutral $M_2$ ($q_1=u$ and $q_2=d$), 
we replace  $Q_{q_1}=Q_{u}=2/3$ and $Q_{q_2}=Q_d=-1/3$ in 
(\ref{eq:HresWI}), in which case 
\begin{equation}
H^{\rm{I}(1)}_{2,0}(u) =  -\frac{1}{54}L^2 +\frac{4}{9}L_\nu-\frac{5}{54}L -\frac{2}{27}g(u) - \frac{\pi^2}{324} +\frac{29}{27}\,.
\end{equation}
We defined
 \begin{align}
     f(u) = {\rm Li}_2(1-u) + 2 \ln^2 u - (3+2i\pi) \ln u -\frac{\ln u }{1-u}
\label{eq:fdef}
 \end{align}
and
\begin{eqnarray}
g(u)& = & 3 \left(\frac{1-2u}{1-u} \ln u - i \pi \right)+ 
\bigg[2 \,{\rm Li}_2(u) - \ln^2 u + \frac{2 \ln{u}}{1-u} -
 (3+2 i\pi)\ln{u} 
\nonumber \\
    && - \,(u \to  1-u ) \bigg]\,.
\label{eq:gdef}
  \end{eqnarray}
The kernels $H_{1,-}^{{\rm I}(1)}(u)$, $H_{1,0}^{{\rm I}(1)}(u)$ 
of the octet operator $Q_1$ follow from the all-order 
identities~(\ref{eq:pureQEDrelations}).

\subsection{\boldmath 
Hard-scattering kernels $H^{\rm {II}\gamma}_{i,\qtwo}$}
\label{subsec:H2}

The logic of the matching calculation for the spectator-scattering 
contribution follows~\cite{Beneke:2006mk}. Here the 
full two-step matching QCD$\times$QED $\to$ SCET$_{\rm I}$
$\to$ SCET$_{\rm II}$ needs to be performed to factorize 
the kernel into its hard and hard-collinear components.
The matching coefficients $H^{\rm {II}\gamma}_{i,\qtwo}(u,v)$ 
in momentum space, which account for the hard contribution 
can be extracted from the on-shell five-point 
$b\to [q_{1}\bar u] \,q_2 \gamma$ amplitude
\begin{equation}
\langle q_1(q_1)\bar u(q_2) q_2(p^\prime_1) \gamma(p_2^\prime)|Q_i|
b(p)\rangle
\label{amp1}
\end{equation} 
and the corresponding SCET${}_{\rm I}$ matrix elements of the
right-hand side of (\ref{eq:matchI}). For the right insertion, 
$q_1=d, q_2=u$ (vice versa for the wrong insertion). 
It is convenient to choose the polarization 
of the external collinear photon state to be transverse to 
$n_\pm^\mu$. With one 
exception, we can set the small transverse components of the 
external momenta to 0, since the operator
${\cal O}_{\qtwo}^{{\rm II}\gamma}$ does not contain transverse 
derivatives for the photon with transverse polarization.
The momenta in the 
anti-collinear direction are then $q_1=u m_B \frac{n_+}{2}$, 
$q_2=\bar u m_B \frac{n_+}{2}$, those of the collinear quark and 
photon are $p_1^\prime =v m_B \frac{n_-}{2}$, 
$p_2^\prime =\bar vm_B \frac{n_-}{2}$, and for the 
heavy quark momentum 
$p^\mu=m_B v^\mu$. For such external momenta the SCET and 
HQET spinors coincide with the QCD ones.

The leading $\mathcal{O}(\alem)$ contributions to the 
$H^{\rm{II}\gamma}_{i,\qtwo}(u,v)$ kernels require only the 
calculation of the tree-level diagrams in Fig.~\ref{fig:diaII}.
Denoting by the angle bracket the matrix element in the 
external state specified above, the matching 
relation (\ref{eq:matchI}) for $Q_2$ amounts to
\begin{equation}
\langle Q_2 \rangle  =   \sum_{i=1}^4 S_i=  
H_{2,\qtwo}^{{\rm I}\gamma, {(\rm tree)}} 
\otimes \langle {\cal O}_{\qtwo}^{{\rm I}}\rangle^{(\rm tree)} + 
H_{2,\qtwo}^{{\rm II}\gamma, {(\rm tree)}} \otimes 
\langle {\cal O}_{\qtwo}^{{\rm II}\gamma} \rangle^{(\rm tree)} \,,
\label{eq:matchexplicit}
\end{equation}
where $\otimes$ denotes the convolution in momentum fractions, 
and $S_i$ the contribution from the four diagrams in 
Fig.~\ref{fig:diaII} ordered as shown. The exception to 
setting the transverse momentum components to 0 applies 
to diagram 4, since the $q_2$-quark propagator with momentum
$p^\prime=p^\prime_1+p^\prime_2$ becomes singular. 
These non-local, long-distance contributions exactly cancel in the 
matching relation against time-ordered products of 
${\cal O}_{\qtwo}^{{\rm I}}$ and the 
SCET interaction Lagrangian \cite{Beneke:2004rc}. 
The local, short-distance 
contribution $S_4{|_{\rm SD}}$ to the matching  
coefficient can be extracted via the 
substitution \cite{Beneke:2006mk}
\be
\frac{i \slashed{p}^\prime}{{p^\prime}^2} \to \frac{i}{n_+ p^\prime}\,
\frac{\slashed{n}_+}{2}\,.
\label{eq:sub1}
\ee
We then find, for the right insertion of $Q_2$,
\begin{align}
S_1 & = 0\,, \nonumber \\
S_2 & = \frac{2 e Q_u}{\bar{u} m_b} \langle  
\frac{\slashed{n}_-}{2} \rangle_{\bar C}  \, 
\bar{\xi}^{(u)}_C \gamma_\perp^\mu (1+\gamma_5) \, 
h_v \epsilon^*_\mu\,, 
\nonumber\\
S_3 & = \frac{2 e Q_d}{m_b} \langle  \frac{\slashed{n}_-}{2} \rangle_{\bar C}  \, \bar{\xi}^{(u)}_C \gamma_\perp^\mu (1+\gamma_5) \, h_v  \epsilon^*_\mu \,,
\nonumber\\
S_4{|_{\rm SD}} & = 0\,,
\label{eq:Siresults}
\end{align}
with abbreviation 
\begin{equation}
  \langle  \frac{\slashed{n}_-}{2} \rangle_{\bar C} \equiv [\bar{\xi}^{(q_1)}_{\bar C} \frac{\slashed{n}_-}{2} (1-\gamma_5) \xi^{(u)}_{\bar C}] \,.
\label{eq:spinorproduct}
\end{equation}
External spinors are denoted by their corresponding fields. 
The advantage of choosing a transversely polarized external photon is 
that with $n_\pm\cdot\epsilon=0$, the tree-level SCET matrix element 
$\langle {\cal O}_{\qtwo}^{{\rm I}}\rangle$ becomes simple. 
Since the collinear photon is decoupled from the 
anti-collinear and heavy-quark fields, the external photon 
can attach only to Wilson lines, which gives zero due to 
$n_-\cdot\epsilon=0$, or to the outgoing collinear quark 
$q_2$. However, the SCET diagram corresponding to the fourth 
diagram $S_4$ in Fig.~\ref{fig:diaII} reproduces the long-distance 
contribution to $S_4$, that was already removed when 
the substitution (\ref{eq:sub1}) was made. 

Hence we set the first term on the right-hand side of 
(\ref{eq:matchexplicit}) to zero. Noticing further 
that the field products 
in (\ref{eq:Siresults}), (\ref{eq:spinorproduct}) match the 
structure of ${\cal O}_{\qtwo}^{{\rm II}\gamma}$, we find 
for the hard-scattering kernels
\begin{equation}
\label{eq:HII}
H^{\rm{II}\gamma}_{2,-}(u,v) = 
N_c \,
H^{\rm{II}\gamma}_{2,0}(u,v) = \frac{2}{\bar{u}} Q_u + 2 Q_d \ , 
\end{equation}
which correspond to the ``right'' and ``wrong'' insertion of 
operator $Q_2$. 
The scattering kernels for $Q_1$ relate to those of $Q_2$ in 
the same way as for $H^{\rm I}_{2,\qtwo}$:
\begin{equation}
    H^{\rm {II} \gamma}_{1,0}(u,v)= C_F H^{\rm {II}\gamma}_{2,0} \,,
\quad \quad   H^{\rm {II} \gamma}_{1,-}(u,v) =0 \ .
\end{equation}
We do not compute QED corrections to the coefficients 
$H^{{\rm II}g}$ of the gluon operators, 
which contribute first at $\mathcal{O}(\alem \alpha_s)$.

We conclude this section with an important remark.
In QCD, only the non-factorizable diagrams $S_{1}+S_2$ 
contribute to the scattering kernels $T^{\rm II}$. 
In this sum, the longitudinally polarized gluons cancel and 
hence it is not necessary to assume the transverse polarization 
for the matching. This is different in QED. When computing these  
diagrams naively by projecting them on the LCDAs of the mesons 
one encounters endpoint divergences that would indicate a 
breakdown of factorization. These arise only from 
longitudinally polarized photons. Consistency of the SCET analysis 
above requires that these terms are contained in the 
matrix element $\langle {\cal O}_{\qtwo}^{{\rm I}}\rangle$ 
of the first operator in the factorization formula, and 
hence they are actually part of the generalized 
heavy-to-light SCET$_{\rm I}$ form factor, which we define below.
We demonstrate this explicitly in Appendix~\ref{app:sec:spec}
by repeating the calculation for general  photon polarization.


\section{SCET$_{\rm I}$ factorization}
\label{sec:SCETIfact}

In~\eqref{eq:opdefscetI} of the previous section we identified the 
SCET$_{\rm I}$ operators relevant to the non-radiative amplitude 
at leading power, discussed their renormalization, and derived 
finite matching coefficients (scattering kernels) at 
$\mathcal{O}(\alem)$. The decoupling of soft photons from the 
anti-collinear sector, which describes the $M_2$ meson, suggests 
that the anti-collinear part (\ref{eq:QEDpiminusLCDA}) of 
these operators and the collinear plus soft part 
(\ref{eq:SCETIbm1op}) (and the corresponding operators with 
an additional hard-collinear photon, see~\eqref{eq:opdefscetI}, 
should be treated as separate entities that renormalize 
independently. However, as already mentioned at the end of 
Section~\ref{sec:QEDERBL}, for the case of an electrically 
charged meson $M_2$, this soft decoupling from $M_2$ does not 
happen, leaving an IR divergent anomalous dimension 
in conflict with the naive SCET factorization. 
Following~\cite{Beneke:2019slt}, we will now see that 
factorization can be restored by a ``soft rearrangement''
that moves a soft overlap contribution between the soft to 
the (anti-) collinear sector.


\subsection{Soft rearrangement}
\label{subsec:softsubtraction}

For charged $M_2$, the UV 
poles~\eqref{eq:UVpolescbar},~\eqref{eq:ZBL} of the 
purely anti-collinear operator~\eqref{eq:QEDpiminusLCDA} depend on 
the IR regulator, in our case the small off-shellness of the 
external partonic momenta. The critical terms originate from the 
soft limit of the anti-collinear propagators in the 
diagrams in Fig.~\ref{fig:QEDERBL}. 
To deal with this soft overlap contribution and make the 
anti-collinear part of the full SCET$_{\rm I}$ operator 
well-defined on its own, we define the rearrangement factors 
$R_c$ and $R_{\bar{c}}$ through
\begin{equation}
\label{eq:Rdef}
\left| \langle 0 |\Big[S_{n_+}^{\dagger (Q_{M_2})} S_{n_-}^{(Q_{M_2})}\Big](0)\,|0\rangle\right| \equiv R_{\bar{c}}^{(Q_{M_2})} R_{c}^{(Q_{M_2})} 
\end{equation}
in close analogy to~\cite{Beneke:2019slt}. 
Taking the absolute value ensures that we do not introduce soft 
rescattering phases into the collinear sector.
We emphasize that the dimensionally regulated 
on-shell vacuum matrix element equals unity to all orders 
in $\alem$, since the soft Wilson lines give rise only 
to scaleless integrals. However, to consistently define the 
renormalized anti-collinear matrix element, we must 
compute~\eqref{eq:Rdef} with the same dimensional UV and 
off-shell IR regularization that was used to 
obtain~\eqref{eq:UVpolescbar},~\eqref{eq:ZBL}. We define the split 
of the vacuum matrix element~\eqref{eq:Rdef} into the two factors 
on the right in such a way that the divergent part of 
$R_{\bar{c}}^{(Q_{M_2})}$ 
depends only on the off-shell regulator 
$\delta_{\bar c}$ in the anti-collinear sector 
defined in (\ref{eq:propmod}), while  $R_{c}^{(Q_{M_2})}$ 
depends only on an accordingly defined $\delta_c$ with 
$n_- \leftrightarrow n_+$. Also the finite terms of these 
factors, which are of no concern in the following, are defined 
such that  $R_{\bar{c}}^{(Q_{M_2})}$ follows from $R_c^{(Q_{M_2})}$ 
through the interchange $n_- \leftrightarrow n_+$.\footnote{
The notation differs from~\cite{Beneke:2019slt}, where the $n_-^\mu$ 
direction is defined as the direction of flight of the negatively 
charged $\ell^-$. In addition, in \cite{Beneke:2019slt} the full 
vacuum matrix element instead of its absolute value was employed
in the definition, and the split into the two factors was made such 
that both had the {\em same} dependence on $\delta_c$ {\em and} 
$\delta_{\bar c}$.} 
We then find 
\begin{equation}
\label{eq:Rcbar}
R_{\bar{c}}^{(Q_{M_2})} = 1 -  \frac{\alem}{4\pi} Q_{M_2}^{\,2} \left[\frac{1}{\epsilon^2} + \frac{2}{\epsilon}\ln{\frac{\mu}{-\delta_{\bar c}}} + \mathcal{O}(\epsilon^0) \right] \ ,
\end{equation}
at $\mathcal{O}(\alem)$. A corresponding expression for 
$R_c^{(Q_{M_2})}$ 
holds by replacing $\delta_{\bar c}\to \delta_c$. 
We assume $\delta_{\bar c}<0$ to not introduce spurious imaginary 
parts from the definition of the IR regulator. 

The rearrangement affects the definition of the QED-generalized 
LCDA for charged light mesons as well as the generalized 
SCET$_{\rm I}$ form factors. 
We redefine the anti-collinear operator~\eqref{eq:QEDpiminusLCDA}
by multiplication with $R_{\bar{c}}^{(Q_{M_2})}$, 
and the QED-generalized LCDA 
and decay constant of the light meson $M_2$ by 
\begin{equation}
\label{eq:M2LCDA}
\langle M_2(p)| R_{\bar{c}}^{(Q_{M_2})} \, 
\bar{\chi}^{(q)}_{\bar{c}}(t n_-) 
\frac{\slashed{n}_-}{2}(1-\gamma_5) \chi^{(u)}_{\bar{c}} (0)|
0\rangle =  \frac{i n_-p}{2}\int_0^1 du\, e^{i u (n_-p) t}  \mathscr{F}_{M_2}\Phi_{M_2}(u)   \ ,
\end{equation}
with $q=u$ or $d$ and $Q_{M_2}=Q_{q}-Q_u$. One can check using 
\eqref{eq:UVpolescbar}, \eqref{eq:deltadef} that \eqref{eq:Rcbar} 
removes the dependence on the IR regulator, which allows to 
renormalize (\ref{eq:M2LCDA}) consistently. The renormalization 
group evolution of this LCDA will not be needed in the following, 
and we defer a detailed discussion to \cite{BLCDApaper}.
When the light meson $M_2$ is neutral, the above definition 
coincides with the standard definition in QCD, since 
$R^{(0)}_{\bar{c}} = 1$ trivially by definition. 

Since the full SCET$_{\rm I}$ operator $\mathcal{O}^{\rm I}_{Q_2}(t)$ 
should not be modified, multiplying its anti-collinear part with 
$R_{\bar{c}}^{(Q_{M_2})}$ requires that we divide the soft and 
collinear 
part by this factor, which defines the generalized SCET$_{\rm I}$ 
$B\to M_1$ form factors as follows: 
\begin{eqnarray}
\label{eq:SCET1FFs}
&& \langle  M_1(p^\prime)| \frac{1}{R_{\bar{c}}^{(Q_{M_2})}} \bar{\chi}^{(q)}_{{C}}(0) \slashed{n}_+(1-\gamma_5) S_{n_+}^{\dagger(Q_{M_2})} h_v (0)|\bar{B}\rangle  =   4 E_{M_1} \zeta^{B M_1}_{Q_2}(E_{M_1})\ , 
\\
&& \langle  M_1(p^\prime)|\frac{1}{R_{\bar{c}}^{(Q_{M_2})}}  \, 
\frac{1}{m_b}\bar{\chi}^{(q)}_{{C}}(0)\frac{\slashed{n}_+}{2} 
\slashed{\cal{A}}_{C,\perp}(sn_+) (1+\gamma_5)
S_{n_+}^{\dagger (Q_{M_2})} h_v (0)|\bar{B}\rangle 
\nonumber\\ 
&& \hspace*{2cm} =\,   
- 2  E_{M_1}\int_0^1 d\tau e^{i\tau (n_+ p^\prime) s}\, 
\Upsilon^{B M_1}_{Q_2}(E_{M_1},\tau)  \ , 
\end{eqnarray}
where $E_{M_1}=n_+ p^\prime/2 = (m_B^2 - q^2)/(2m_B)$ is the energy 
of meson $M_1$ in the $B$-meson rest frame for vanishing 
light-meson mass $m_{M_1} = 0$. 
For the matrix element of the gluonic operator 
$\mathcal{O}_{Q_2}^{{\rm II}g}$ we replace 
$\cal{A}\to \cal{G}$ and $\Upsilon^{B M_1}_{Q_2}(E_{M_1},\tau)
\to  \Sigma^{B M_1}_{Q_2}(E_{M_1}, \tau)$.
Since the full SCET$_{\rm I}$ operator and the anti-collinear 
operator after the soft rearrangement are well-defined, so are 
the generalized form factors. Note that they carry information 
about the meson $M_2$, but only of its charge  $Q_{M_2}$ and 
direction of flight $n_+^\mu$ through the additional Wilson line 
$S_{n_+}^{\dagger (Q_{M_2})}$ and soft rearrangement,  
as expected from the universality of 
soft interactions. The definitions above are such that in the 
pure-QCD limit $\alem\to 0$ the form factors $(\zeta,\Sigma)$ 
reduce to $(\xi,\Xi)$  in the notation 
of~\cite{Beneke:2005vv}, and $\Upsilon\to 0$.

With these preparations, taking matrix elements of (\ref{eq:matchI}), 
the SCET$_{\rm{I}}$ factorization formula reads 
\begin{eqnarray}
\label{eq:SCET1Fac}
\bra{M_1 M_2}Q_i \ket{\bar B} 
&=& \, i m_B^2 \bigg\{ 
\,\zeta_{Q_2}^{B M_1}  \int_0^1 du \, {H}^{{\rm I}}_{i,\qtwo}(u) \mathscr{F}_{M_2}\Phi_{M_2}(u)  
\nonumber \\
&& \hspace*{-2.5cm}- \,\frac{1}{2} \, \int_0^1 du \; dz 
  \Big[ H^{{\rm II}\gamma}_{i,\qtwo}(u,z)\,
\Upsilon_{Q_2}^{B M_1}({1-z})  + H^{{\rm II}g}_{i,\qtwo}(u,z) 
\,\Sigma^{B M_1}_{Q_2}({1-z}) \Big] \mathscr{F}_{M_2}\Phi_{M_2}(u) \,
\bigg\} \,,\qquad
\end{eqnarray}
where we have dropped the energy argument of the form factors, 
which is $E_{M_1}=m_B/2$ here. 


\subsection{The soft form factor and the 
semi-leptonic amplitude}
\label{sec:softff}

In the first line of (\ref{eq:SCET1Fac}) we recognize the 
first line of the previously stated QED factorization 
formula \eqref{eq:QEDF}, if we identify
\be
\mathcal{F}^{BM_1}_{Q_2}(q^2=0)\to 
\zeta_{Q_2}^{B M_1}(E_{M_1}=m_B/2), 
\qquad T^{{\rm I}}_{i,Q_{2}}(u) \to  {H}^{{\rm I}}_{i,\qtwo}(u)\,.
\ee
In pure QCD, at this point one replaces the cooresponding 
SCET$_{\rm I}$ form factor $\xi^{B M_1}(E)$ by the full QCD form factor, using a
similar factorization formula for the form factor 
\cite{Beneke:2000wa}, since it is the full QCD form factors  
which are calculated with light-cone QCD sum rules or 
lattice QCD. The factorization formula~\eqref{eq:SCET1Fac} including 
QED effects contains the QED-generalized light-meson LCDA and the 
generalized SCET$_{\rm I}$ form factor $\zeta_{Q_2}^{B M_1}(E)$. 
If $M_2$ is neutral, the latter can again be replaced by 
the full QCD$\times$QED theory matrix element of the local 
heavy-to-light 
current operator, which corresponds to the usual form factor, 
but including QED effects. However, for charged $M_2$, 
the generalized form factor $\zeta_{Q_2}^{B M_1}(E)$ contains a 
Wilson line that knows about $M_2$, which cannot be written 
as the matrix element of a local operator. The physical quantity 
in the full theory with the same IR physics is now the 
non-radiative amplitude of the semileptonic decay  
$\bar B \to M_1 \ell^- \bar{\nu}_{\ell}$ in 
the kinematic point $q^2 = 0$ and $E_{\ell} = m_B/2$. 
We therefore consider eliminating  $\zeta_{Q_2}^{B M_1}(E)$ 
in favour of this semi-leptonic amplitude by making 
use of the QED generalization of the factorization 
theorem for heavy-to-light form factors, discussed in 
generality in \cite{semileptonicpaper}.

\subsubsection{Semi-leptonic QED factorization}

The Hamiltonian for the $b \to u \ell^- \bar{\nu}_\ell$ transition is 
\begin{equation}
      \mathcal{H}_{\rm sl} = \frac{G_F}{ \sqrt{2} } V_{ub} C_{\rm sl} Q_{\rm sl}\ ,
\end{equation}
with
\begin{equation}
    Q_{\rm sl}=\bar{u} \gamma^\mu (1-\gamma_5) b \, \bar{\ell} \gamma_\mu (1-\gamma_5) \nu \ .
\end{equation}
Unlike in pure QCD, where $C_{\rm sl}=1$ to all orders in the 
strong coupling, the semi-leptonic $b\to u \ell^- \bar{\nu}_\ell$ 
transition receives short-distance QED and electroweak corrections 
from the scale  $\mathcal{O}(m_W)$. The Wilson coefficient  
$C_{\rm sl} = C_{\rm sl}(\nu)$ evolves under the renormalization 
group from $m_W$ to the scale 
$\nu\sim \mathcal{O}(m_b)$,
which sums large logarithms $\alem^n
\ln^m (m_W/m_b)$ ($m \leq n$).\footnote{The Fermi constant $G_F$, 
defined as the short-distance $\mu^- \to e^- \nu_\mu \bar{\nu}_e$ 
decay amplitude (that is, excepting low-energy QED corrections), 
is not renormalized.}
The product $C_{\rm sl} \, Q_{\rm sl}$ is independent of the 
scale $\nu$. 
In essence, as far as electroweak and QED effects are concerned, 
$Q_{\rm sl}$ is not very different from the four-quark operators.
The one-loop expression for the Wilson coefficient is 
known from \cite{Sirlin:1981ie}, see also (\ref{eq:csl}) below. 
The non-radiative semi-leptonic amplitude is given by 
\begin{align}
\mathcal{A}^{{\rm sl},M_1}_{\text{non-rad}}& = 
\frac{G_F}{\sqrt{2}} V_{ub} C_{\rm sl}\,
\bra{M_1 \ell^- \bar{\nu}_\ell}Q_{\rm sl} \ket{\bar B} \nonumber \\
& \equiv \frac{G_F}{\sqrt{2}} V_{ub}\,  4 E_{M_1} [\bar{u}(p_\ell) 
\,\frac{\slashed{n}_-}{2} (1-\gamma_5) v_{\nu_{\ell}}(p_\nu)] 
\,\mathcal{A}^{{\rm sl},M_1}_{\rm red}\,.
\end{align}
The second line defines the $\nu$-independent 
reduced amplitude $\mathcal{A}^{{\rm sl},M_1}_{\rm red}$, 
which is the analogue of the standard form factor for the 
case of the electrically neutral $M_2$.

A completely analogous analysis of 
QCD$\times$QED $\to$ SCET$_{\rm I}$ matching for the 
semi-leptonic operator $Q_{\rm sl}$ instead of the hadronic 
operators $Q_{1,2}$ results in 
\begin{eqnarray}
\label{eq:SCET1Facsemilep} 
\bra{M_1 \ell^- \bar{\nu}_\ell}Q_{\rm sl} \ket{B} 
  &=& \, 4 E_{M_1} [\bar{u}(p_\ell) \frac{\slashed{n}_-}{2} (1-\gamma_5) v_{\nu_{\ell}}(p_\nu)] \, Z_{\ell} \, \bigg\{ H_{\rm sl}^{\rm I}(E_{\ell})\,\zeta_{-}^{B M_1}(E_{M_1})   \,  \nonumber \\
 && 
\hspace*{-2.5cm}  -  \,\frac{1}{2} \, \int_0^1 dz 
  \Big[H_{\rm sl}^{{\rm II}\gamma}(E_{\ell},z) 
\,\Upsilon_{-}^{B  M_1}(E_{M_1},{1-z}) 
+ H_{\rm sl}^{{\rm II}g}(E_{\ell},z) 
\Sigma^{B M_1}_{-}(E_{M_1},{1-z}) \Big]  \bigg\},\qquad
\end{eqnarray}
which can be compared to \eqref{eq:SCET1Fac} for $Q_2=-$, where 
the same generalized form factors appear. That the lepton 
$\ell^-$ is point-like entails some simplifications. Instead of 
the LCDA of $M_2$ defined through the matrix element of 
(\ref{eq:QEDpiminusLCDA}), we need the matrix element of 
the anti-collinear point-like lepton field 
${\chi}^{(\ell)}_{\bar C}=[W_{\bar C}^{(\ell)}]^\dagger 
\xi_{\bar C}^{(\ell)}$ 
\cite{Beneke:2019slt}, which defines the factor $Z_\ell$ 
in (\ref{eq:SCET1Facsemilep}), and there is no integral over $u$. 

The UV renormalization of $Z_\ell$ follows the same line of 
reasoning as for the light-meson LCDA, but is technically 
simpler. The dimensionally UV and off-shell IR regulated 
matrix element of the dressed lepton field operator is 
\begin{equation}
\bra{\ell^-(p_\ell)} \bar{\chi}^{(\ell)}_{\bar{C}}(0) \ket{0} =
\bar{u}(p_\ell) \,\frac{\slashed{n}_-\slashed{n}_+}{4} 
\left\{ 1 + \frac{\alem}{4\pi} Q_\ell^2 \left[ \frac{2}{\eps^2} 
+ \frac{3}{2\eps} +
\frac{2}{\eps} \log \frac{\mu^2}{-p_\ell^2} + \mathcal{O}(\eps^0) 
\right] \right\} \,.
\end{equation}
The UV pole depends on the IR regulator $p_\ell^2$, as was 
to be expected, since we must multiply with the soft 
rearrangement factor 
\begin{equation}
R_{\bar{c}}^{(Q_\ell)} = 
1 - \frac{\alem}{4\pi} Q_\ell^2 
\left[ \frac{1}{\eps^2} + \frac{2}{\eps} \ln 
\frac{\mu}{-\delta_{\bar c}} +
\mathcal{O}(\eps^0)\right] = R_{\bar{c}}^{(Q_{M_2})} \,,
\end{equation}
which was divided out in defining the generalized form factors 
appearing in (\ref{eq:SCET1Facsemilep}). Consistency requires 
that the same $\delta_{\bar c}$ appears here, so that the 
second equality holds, that is, $p_\ell^2$ 
must be chosen such that $p_\ell^2/n_-p_\ell = \delta_{\bar c}$ 
with $n_-p_\ell = 2E_\ell$.
Similar to (\ref{eq:M2LCDA}) for the soft-rearranged LCDA for 
the light meson, we now define $Z_\ell$ for the point-like lepton 
via
\begin{equation}
\label{eq:Zldef}
\bra{\ell^-(p_{\ell}) } R_{\bar{c}}^{(Q_{\ell})} \bar{\chi}^{(\ell)}_{\bar C}(0)  \ket{0} 
    \equiv Z_{\ell} \bar{u}(p_\ell)\frac{\slashed{n}_- \slashed{n}_+}{4}\,,
\end{equation}
and obtain 
\begin{equation}
Z_{\ell}^{\rm bare} = 
1 + \frac{\alem}{4\pi} Q_\ell^2 \left[ \frac{1}{\eps^2} + \frac{3}{2\eps} + \frac{2}{\eps} \ln \left( \frac{\mu}{n_- p_\ell}
\right) + \mathcal{O}(\eps^0) \right].
\label{eq:Zbare}
\end{equation}
The given pole part determines the UV renormalization constant, 
which is the analogue of (\ref{eq:ZBL}), but for the case of a 
point-like charged particle. The quantity $Z_\ell$ that enters 
(\ref{eq:SCET1Facsemilep}) above is the UV renormalized on-shell 
matrix element $\bra{\ell^-(p_{\ell}) } R_{\bar{c}}^{(Q_{\ell})} \bar{\chi}^{(\ell)}_{\bar C}(0)  \ket{0}$, see \eqref{eq:Zlonshell} 
below.

The matching coefficients 
$H_{\rm sl}^{\rm I}(E_{\ell})$ and $H_{\rm sl}^{\rm II}(E_{\ell},z)$ 
in (\ref{eq:SCET1Facsemilep}) can be obtained from the general
 expression~\eqref{eq:Hres} for 
$H^{\rm I}(u)$ and~\eqref{eq:HII} for $H^{\rm II}$ by replacing 
$Q_{q_1}\to Q_\ell, Q_u\to 0$ and $u \to 2E_{\ell}/m_B$. 
Setting now $E_{\ell} = m_B/2$ (for this value we drop the 
lepton-energy argument of the matching coefficients), we 
find  $H_{\rm sl}^{\rm{I}(0)}=1$ and  
\begin{eqnarray}
&& H_{\rm sl}^{\rm{I}(1)} = 
 Q_\ell Q_u \left(L^2 -3L_\nu+\left(3+ 2i\pi\right) L-  \frac{7\pi^2}{6}-2\right) - Q_d^2 \left(\frac{1}{2}L^2 + \frac{5}{2}L + \frac{\pi^2}{12} +6\right), \qquad
\label{eq:HslI}\\
&& H_{\rm sl}^{\rm{II}\gamma}(z) = 2  Q_d  \,.
\label{eq:HslII}
\end{eqnarray}

\subsubsection{\boldmath
Introducing $\mathcal{A}^{{\rm sl},M_1}_{\rm red}$}

For charged $M_2$, we now use the factorization formula for the 
reduced semi-leptonic amplitude 
$\mathcal{A}^{{\rm sl},M_1}_{\rm red}$ 
implied by \eqref{eq:SCET1Facsemilep}  to eliminate
the SCET$_{\rm 1}$ form factor $\zeta_{-}^{B M_1}$. For neutral 
$M_2$, we follow the standard QCD procedure \cite{Beneke:2005gs} 
and replace  $\zeta_{0}^{B M_1}$ 
by the full QCD$\times$QED $B\to M_1$ transition form factor. 
This gives the factorization formula
\begin{eqnarray}
\label{eq:facIchar}
\left\langle M_1 M_2 | Q_i |\bar B\right\rangle 
&=& im_B^2 \,\bigg\{  \mathcal{F}_{Q_2}^{BM_1}(0)\, \int_0^1 du \, T_{i, \qtwo}^{\rm I}(u) \, \mathscr{F}_{M_2} \Phi_{M_2}(u) 
\nonumber \\
&& \hspace*{-2.5cm}
-\,\frac12 \int_0^1 du \;dz \, \left[ 
\hat{H}_{i,\qtwo}^{\rm{II}\gamma}(u,z)  
\Upsilon_{Q_2}^{BM_1}(1-z) + 
\hat{H}_{i,\qtwo}^{\rm{II}g}(u,z)  
\Sigma_{Q_2}^{BM_1}(1-z)\right] 
\mathscr{F}_{M_2} \Phi_{M_2}(u) \bigg\} \,, \qquad
\end{eqnarray}
where now 
\begin{equation}
\mathcal{F}^{BM_1}_{-}(0) \equiv 
\frac{\mathcal{A}^{{\rm sl},M_1}_{\rm red}}{C_{\rm sl} Z_{\ell}}
\label{eq:semilepFF}
\end{equation} 
in the form-factor term in the first line is expressed in 
terms of the reduced semi-leptonic amplitude at $q^2=0$, while 
$\mathcal{F}^{BM_1}_{0}(0)$  are the full QCD$\times$QED 
$B\to M_1$ transition form factors as in pure QCD.
The hard-scattering kernels also change and are now given by
\begin{equation}
T_{i,-}^{\rm I}(u; E_{\ell})\equiv \frac{H_{i,-}^{\rm I}(u)}{H_{\rm sl}^{\rm I}(E_{\ell})} \ , \quad \quad\quad  T_{i,0}^{\rm I}(u)\equiv \frac{H_{i,0}^{\rm I}(u)}{{H}_{f}^{\rm I}} \,,
\label{eq:TIkernels}
\end{equation}
and
\begin{eqnarray}
\label{eq:HIInonfac2}
&& \hat{H}^{\rm II \gamma}_{2,-}(u,z;E_\ell) =
H^{\rm II \gamma}_{2,-}(u,z) - T_{2,-}^{\rm I}(u;E_\ell) 
H_{\rm sl}^{\rm II\gamma}(z) = \frac{2}{\bar{u}}Q_u \,,
\label{eq:HIInonfac}\\
&& \hat{H}^{\rm II \gamma}_{1,0}(u,z) =
C_F \hat{H}^{\rm II \gamma}_{2,0}(u,z)=  H_{1,0}^{\rm II \gamma}(u,z) - T_{1,0}^{\rm I}(u) H_f^{\rm II\gamma}(z) 
=\frac{2 C_F}{N_c \bar{u}}Q_u \ ,
\label{eq:HIInonfac1}
\end{eqnarray}
while $\hat{H}^{\rm II \gamma}_{1,-}(u,z)$ remains zero.
$H_{f}^{\rm I(1)}$ in (\ref{eq:TIkernels}) and 
${H}_{f}^{\rm II\gamma}(z)$ are the matching coefficients 
in the SCET$_{\rm I}$ factorization formula of the  
full QCD$\times$QED transition form factor.\footnote{This definition differs from the factorization formula in \cite{Beneke:2005gs} by a factor of $-\frac{1}{2}$ for the spectator-scattering terms, therefore, in the QCD case ($Q_d\to 1$) our coefficient $H_f^{\rm{II}\gamma} \to -2C_{f_+}^{(B1)}$, where the latter is the QCD coefficient defined in \cite{Beneke:2005gs}.} They can be  obtained from 
the semi-leptonic coefficients (\ref{eq:HslI}), (\ref{eq:HslII})
by putting $Q_\ell=0$:
\begin{eqnarray}
    &&{H}_{f}^{\rm I(1)}  =- Q_d^2 \left[\frac{1}{2}L^2 + \frac{5}{2} L+ \frac{\pi^2}{12} + 6 \right],\\
    &&{H}_{f}^{\rm II\gamma (0)}(z)  = 2 Q_d \, .
\end{eqnarray} 

We remark that the normalization to the semi-leptonic amplitude 
requires only $q^2=0$, but any value of the lepton energy $E_\ell$ 
 can be used as long as it is $\mathcal{O}(m_B/2)$. Then, the
scattering kernels $T_{i,-}^{\rm I}(u;E_\ell)$ and 
$\hat{H}^{\rm II \gamma}_{2,-}(u,z;E_\ell)$ acquire a  
dependence on $E_\ell$ as indicated by their additional 
argument. We dropped this argument in (\ref{eq:facIchar}). 
For simplicity, we give here the results for the kernels 
for $E_{\ell}=m_B/2$, as the general result can be easily 
obtained from the above results. We find
\begin{eqnarray}
\label{eq:Tall}
T^{\rm{I}(1)}_{1,-}(u) & =& 0\,, \nonumber\\
T^{\rm{I}(1)}_{2,-}(u) & = &-\frac{2}{3}L_\nu +\frac{2}{3}L
\left( 2\ln(1-u) + \ln u\right) \nonumber \\
     &&- \frac{2 f(u) + 4f(1-u)}{9}- \frac{(2-u) \ln(u)}{3(1-u)}   -\frac{4i\pi}{3}-\frac{25}{9}\,,\nonumber \\
T^{\rm{I}(1)}_{1,0}(u) & =& C_F T^{\rm{I}(1)}_{2,0}(u) 
=\frac{16}{27}L_\nu -\frac{8}{81}g(u) + \frac{140}{81}\,,
\end{eqnarray}
with $f(u)$, $g(u)$ defined in \eqref{eq:fdef}, \eqref{eq:gdef}, 
respectively. At tree level, there is no change, and 
$T^{\rm{I}(0)}_{i,\qtwo}(u) = H^{\rm{I}(0)}_{i,\qtwo}(u)$.
We note that the double-logarithmic $L^2$ 
terms present in $H^{\rm{I}(1)}_{i,\qtwo}$ have disappeared after 
introducing the semi-leptonic amplitude or full-theory form 
factors. The $L_\nu$ terms are related to the renormalization 
of the operators $Q_i$, $Q_{\rm sl}$, 
and the dependence on the scale $\nu$ cancels with 
the $\nu$ dependence of $C_i(\nu)$ and $C_{\rm sl}(\nu)$. 
The left-over single logarithm of $L$ in 
$T^{\rm{I}(1)}_{2,-}(u)$ appears, because unlike the light-meson 
decay constant in QCD, the QED-generalized 
decay constant $\mathscr{F}_{M_2}$ of a charged meson 
is scale-dependent. The $\mu$ dependence of the one-loop 
kernel is related to the UV divergence of the one-loop bare 
hadronic matrix element convoluted with the tree-level 
kernels. From (\ref{eq:UVpolescbar}), (\ref{eq:Rcbar}) and 
(\ref{eq:Zbare}) (which enters through (\ref{eq:semilepFF})), 
we obtain 
\be
R_{\bar c}^{(Q_{M_2})(1)} +  
\frac{2}{\epsilon} \, \int_0^1 dv \, V(u,v)|_{Q_{q_1}=Q_d}
-Z_\ell^{(1)}
= -\frac{2}{\epsilon}Q_{M_2}\left[
Q_d \ln u-Q_u \ln (1-u)\right]\,,
\label{eq:polecancellation}
\ee
in agreement with the coefficient of $L$ in (\ref{eq:Tall}). 
In general, the 
$\mu$-scale dependence of $T^{\rm{I}(1)}_{2,-}(u)$ cancels 
against $\mathscr{F}_{M_2} \Phi_{M_2}(u)/Z_\ell$ under the 
convolution in (\ref{eq:facIchar}).


\section{SCET$_{\rm II}$ factorization}
\label{sec:spectatorscat}

In the case of pure QCD, the SCET$_{\rm I}$ operators of the 
$\mathcal{O}^{{\rm II}g}_{Q_2}$ type are further matched to 
four-fermion operators in SCET$_{\rm II}$, and the 
corresponding generalized $B\to M_1$ form  factor 
$\Sigma_{\qtwo}^{BM_1}(E_{M_1},1-z)$ is 
expressed in terms of the convolution of a hard-collinear 
matching coefficient with the $B$-meson and light-meson 
LCDAs. This results in the standard form of the 
spectator-scattering term in the QCD factorization 
formula for non-leptonic $B$ decays. 
This can be done, because it can be shown 
\cite{Beneke:2003pa} that these convolutions are convergent 
to all orders in perturbation theory, which has been 
confirmed explicitly by one-loop 
calculations~\cite{Beneke:2005vv, Beneke:2005gs}. 

\subsection{\boldmath Generalized $B$-meson LCDA}

The same matching applies to $\mathcal{O}^{{\rm II}\gamma}_{Q_2}$ and 
$\mathcal{O}^{{\rm II}g}_{Q_2}$ with QED included, but the 
LCDAs have to be appropriately generalized. The definition of the 
LCDA of $M_1$ is analogous to that of $M_2$ in~\eqref{eq:M2LCDA} 
with obvious replacements of anti-collinear and collinear, and $n_-$ 
by $n_+$, as well as $R_{\bar{c}}\to R_c$ to rearrange the 
soft overlap between the collinear and the soft sector. As concerns 
the $B$-meson LCDA, in QCD$\times$QED we must 
distinguish between the charged $\bar{B}_u = B^-$ and neutral 
$\bar{B}_d^0$, $\bar{B}_s^0$ mesons. Since the $B$-meson LCDA is 
the soft function of the process, which inherits the soft Wilson 
lines from the decoupling of the anti-collinear and collinear 
sector,\footnote{While in the first factorization step, we performed 
the decoupling only from the anti-collinear sector, in the 
present matching to SCET$_{\rm II}$, we must finally also 
perform the soft-decoupling field redefinition of the collinear 
fields.}
the electric charges of the 
emitted mesons $M_1$ and $M_2$ also matter, 
leading to a total of four different $B$ LCDAs, defined as
\begin{eqnarray}
\label{eq:BLCDAdef}
&& im_{B} \int_{-\infty}^\infty d\omega \;e^{-i \omega t} \mathscr{F}_{B,\otimes} \Phi_{B, \qboth}(\omega)\nonumber \\[-0.2cm]
&& \hspace*{1cm} = \,
\frac{1}{R_c^{(Q_{M_1})}R_{\bar{c}}^{(Q_{M_2})}} \bra{0} 
\bar{q}^{(q)}_s(tn_-) [tn_-,0]^{(q)}  
\,\slashed{n}_-\gamma_5 h_v(0) S_{n_+}^{\dagger (Q_{M_2})} 
S_{n_-}^{\dagger, (Q_{M_1})} \ket{\bar{B}}  \,. 
\qquad
\end{eqnarray}
The matrix element is divided by the $R_{c}$, $R_{\bar{c}}$ factors 
to compensate their multiplication of the  $M_1$, $M_2$ LCDA, 
and hence depend on the meson charges $Q_{M_1}$, $Q_{M_2}$. While 
the definition looks familiar to pure QCD definition 
with respect to the finite-distance Wilson line $[tn_-,0]^{(q)}$, 
the addition of the Wilson line $S_{n_+}^{\dagger (Q_{M_2})}$ 
in the anti-collinear direction 
leads to fundamentally different properties. For example, this 
$B$-meson LCDA includes the physics of soft rescattering, 
including phases. It might be more useful to think of it as 
the soft function for the $B\to M_1 M_2$ process rather than 
a LCDA. A technical manifestation of this difference is that, 
for charged $M_2$,  
the ``LCDA'' $\mathscr{F}_{B,\otimes} \Phi_{B, \qboth}(\omega)$ 
has support not only for $\omega >0$ but also for negative 
$\omega$ as indicated by the lower limit of the integral. 
We discuss this and the renormalization of these new objects 
in \cite{BLCDApaper}. 


\subsection{Spectator scattering and 
complete factorization}
The matching equation from SCET$_{\rm I} \to$ SCET$_{\rm II}$ 
is \cite{Beneke:2005vv, Beneke:2005gs}  
\begin{equation}
\Upsilon^{BM_1}_{Q_2}(1-z) = \frac{1}{4}\int_{-\infty}^\infty 
d\omega \int_0^1 dv \,J_{\qboth}(1-z;v,\omega)  \mathscr{F}_{B,\qboth} \Phi_{B,\qboth}(\omega)  \mathscr{F}_{M_1} \Phi_{M_1}(v) \,,
\label{eq:Upsilonfact}
\end{equation}
which defines the hard-collinear matching coefficient 
(``jet'' function) $J_\qboth(z;v,\omega)$. 

Tree-level matching gives 
\begin{equation}\label{eq:jetfun}
J_\qboth(\bar{z};v,\omega) = - \frac{4\pi \alem
Q_{\rm sp}}{N_c}\frac{1}{ m_B \omega\bar{v}} 
\,\delta(\bar{z}-\bar{v}) \ ,
\end{equation}
with $Q_{\rm sp} = Q_d - Q_{M_1} -Q_{M_2}$ the charge of the 
spectator-quark $q$ in the $\bar{B}_q$ meson. Inserting 
(\ref{eq:Upsilonfact}) into \eqref{eq:facIchar} 
gives\footnote{Since we focus on QED effects, we omit the 
spectator-scattering contribution $\Sigma^{BM_1}_{\qtwo}(1-z)$ 
from the gluonic operator 
$\mathcal{O}^{{\rm II}g}_{Q_2}$. QED corrections to this term 
are $\mathcal{O}(\alem\alpha_s)$, beyond the accuracy 
of the present work.} 
\begin{eqnarray}
\label{eq:QEDFagain}
\left\langle M_1 M_2 | Q_i |\bar B\right\rangle &=& i m_B^2\,
\bigg\{
\mathcal{F}^{BM_1}_{Q_2}(0) \, \int_0^1 du \, 
T^{{\rm I}}_{i,Q_{2}}(u) \mathscr{F}_{M_2} \Phi_{M_2}(u) 
\nonumber \\
&&\hspace*{-1.5cm}+ 
\,\int_{-\infty}^\infty d\omega \int_0^1 du \;dv \,
T^{{\rm II}}_{i, \qboth}(u,v,\omega) 
\mathscr{F}_{M_1} \Phi_{M_1}(v) \mathscr{F}_{M_2}\Phi_{M_2}(u) 
\mathscr{F}_{B, \qboth} \Phi_{B,\qboth} (\omega)
\,\bigg\} \,,\quad
\end{eqnarray}
which is (\ref{eq:QEDF}). In the spectator-scattering term in the 
second line, the SCET$_{\rm I}$ hard-scattering kernel 
$H_{i,\qtwo}^{\rm II\gamma}$ is convoluted with the 
jet function $J_\qboth$, defining 
\begin{equation}
T^{\rm II}_{i,\qboth}(\omega,u,v)=- \frac{1}{8}\int_0^1 dz 
\,\hat{H}^{\rm II}_{i,\qtwo}(u,z) J_{\qboth}
(1-z;v,\omega) \,.
\end{equation}
Combining (\ref{eq:HIInonfac2}),  (\ref{eq:HIInonfac1})
with ~\eqref{eq:jetfun} gives 
\begin{equation}
    T^{\rm{II}}_{2,(Q_1,-)}(\omega, u, v)=N_c  T^{\rm{II}}_{2,(Q_1,0)} = \frac{N_c}{C_F}T^{\rm{II}}_{1,(Q_1,0)} =\frac{\pi \alem 
Q_{\rm sp} Q_u}{N_c}\frac{1}{m_B \omega\bar{u}\bar{v}} \,
\label{eq:Trelations}
\end{equation}
and $T^{\rm{II}}_{1,(Q_1,-)}(\omega, u, v)=0$ at 
$\mathcal{O}(\alem)$. This completes the factorization of 
QED effects for the matrix elements 
$\left\langle M_1 M_2 | Q_i |\bar B\right\rangle$.


At this point it is worth recalling that the factorization 
discussed so far refers to the non-radiative amplitude, i.e. the 
purely virtual corrections. 
Such non-radiative amplitudes are IR divergent 
for all decays that involve charged mesons. Real emission 
of soft photons must be added to obtain an observable, 
as will be done in the following section.
The theoretical approach developed here applies when the energy of the real photons is much smaller than $\Lambda_{\rm QCD}$, such that the hard and hard-collinear propagators are not affected and the 
corresponding coefficient functions  
are from virtual corrections only.

After integrating out the hard and hard-collinear scales, 
the IR singularities of the non-radiative amplitude are 
hidden in the hadronic matrix elements (soft form factors, heavy and 
light meson LCDAs) of SCET operators, which are all defined 
as non-radiative quantities. The concept of non-perturbative but 
IR divergent LCDAs appears counter-intuitive. 
However, the hadronic scale $\Lambda_{\rm QCD}$ does not 
necessarily act as a regulator for soft IR singularities in QED. 
These hadronic matrix elements should themselves be 
considered as short-distance matching coefficients, when 
SCET$_{\rm II}$ is matched to a very low-energy theory of 
point-like mesons coupled to photons with energy below  
$\Lambda_{\rm QCD}$. In this matching the IR divergence of 
the hadronic matrix elements is removed, but leaves 
a dependence on the IR factorization scale $\mu_{\rm IR}$, 
where this matching is performed. This must be distinguished 
from their UV renormalization scale ($\mu$) dependence, 
which was computed above,  
that follows from the UV poles in dimensional regularization, 
when the corresponding partonic matrix elements are 
computed with an off-shell IR regulator. While the $\mu$ 
dependence can be calculated perturbatively, the 
IR matching of SCET$_{\rm II}$ to the theory of point-like 
mesons must be done {\em non-perturbatively} at a scale a 
few times smaller than $\Lambda_{\rm QCD}$.

We illustrate these points with the help of the UV renormalized leptonic collinear matrix element defined in~\eqref{eq:Zldef}, which can reliably be computed in perturbation theory, since QCD does not 
enter (modulo photon vacuum polarization etc. in higher orders).
In fact, $Z_\ell$ is the weakly-interacting point-particle analogue 
of the LCDA for a strongly interacting composite hadron.
At $\mathcal{O}{(\alem)}$ we compute the single contributing 
on-shell one-loop diagram, add the on-shell renormalization 
factor~\eqref{eq:OSZ} (replacing $m_B\to m_\ell$ and 
$Q_d\to Q_\ell$) and the UV counterterm given by minus 
the divergent part of~\eqref{eq:Zbare}, and 
obtain for the UV renormalized on-shell matrix element
\begin{eqnarray}
Z^{(1)}_\ell 
    &= & - \frac{1}{\eps_{\rm IR}} \left(1 + \ln \frac{m_\ell^2}{m_B^2} \right)  + \frac12 \ln \frac{\mu^2}{m_\ell^2} + \frac12 \ln^2 \frac{\mu^2}{m_\ell^2} + 2 + \frac{\pi^2}{12}\\
    &=& - \left(\frac{1}{\eps_{\rm IR}} + \ln \frac{\mu_{\rm IR}^2}{m_\ell^2} \right) \left(1 + \ln \frac{m_\ell^2}{m_B^2} \right) + 
\frac{3}{2} \ln \frac{\mu^2_{\rm UV}}{m_\ell^2} + \frac12 \ln^2 \frac{\mu_{\rm UV}^2}{m_B^2} - \frac12 \ln^2 \frac{m_\ell^2}{m_B^2} + 2 + \frac{\pi^2}{12}\,.
 \nonumber 
\label{eq:Zlonshell}
\end{eqnarray}
Here $m_\ell$ is the lepton mass, which must be kept at the collinear 
scale, and provides a physical cut-off of the collinear singularities.
Since we subtracted the UV poles, the $1/\eps$ pole must be an IR 
singularity. It is cancelled after matching onto the theory 
of point-like objects (here the lepton itself), where the large logarithm in the ratio $m_\ell/m_B$ arises from the large relative boost between the rest frames of the external particles.
In the second line we use $\mu_{\rm UV}=\mu_{\rm IR}=\mu$ to 
separate the UV and the IR scale dependence.
The UV scale dependence is dictated by the UV poles 
of~\eqref{eq:Zbare}, and is cancelled against the scale dependence of the hard-scattering kernel and the light-meson LCDA, see discussion 
around~\eqref{eq:polecancellation}.
On the other hand, the $\mu_{\rm IR}$ dependence is associated with the ultrasoft function as will be seen in the next section.


\section{Ultrasoft photons and decay rates}
\label{sec:usoft}

So far we studied the non-radiative amplitude for the purely 
exclusive process $B \to M_1 M_2$. Any IR finite observable must 
account for final states with photons of arbitrarily small 
energy, once $M_1 M_2$ contains electrically charged 
mesons.\footnote{The non-radiative amplitude was computed 
setting the light-meson masses to zero, which is justified 
for the computation of the hard and hard-collinear matching 
coefficents, which involve scales far above the meson 
masses. In the ultrasoft theory discussed in this section 
the light-meson masses must be kept, hence there are no 
collinear singularities.} A physically meaningful observable is 
the soft-photon-inclusive decay rate
\begin{equation}
\Gamma[\bar{B} \to M_1 M_2](\Delta E) \equiv \Gamma[\bar{B} 
\to M_1 M_2 + X_s]\big\vert_{E_{X_s} \leq \Delta E}\,,
\end{equation}
where the final state $X_s$ consists of photons and possibly also 
electron-positron pairs with total energy less than $\Delta E$ in the 
$B$-meson rest frame. In the following we assume that 
$\Delta E \ll m_{M_i} \sim \Lambda_{\rm QCD}$ and 
refer to the scale $\Delta E$ as ``ultrasoft'' to distinguish 
it from the soft scale $\Lambda_{\rm QCD}$ relevant to the 
generalized $B$-meson LCDA. 

The $B \to M_1 M_2 + X_s$ amplitude factorizes into the 
non-radiative amplitude discussed before and an ultrasoft 
matrix element. Up to corrections of
$\mathcal{O}(\Delta E/\Lambda_{\rm QCD})$, 
\begin{align}
\label{eq:usoftfac}
\mathcal{A}(\bar{B} \to M_1 M_2 + X_s) &= 
\mathcal{A}(\bar{B} \to M_1 M_2) \, \langle X_s | (\bar{S}_v^{(Q_{B})} S^{\dagger (Q_{M_1})}_{v_1} S^{\dagger (Q_{M_2})}_{v_2})(0) | 0 \rangle \,,
\end{align}
where the $S^{(Q_{M_i})}_{v_i}$ are outgoing time-like Wilson lines, 
defined in analogy to~\eqref{eq:softwilsonlines}, but with 
velocity labels $v_i$ of meson $M_i$, 
satisfying $v_i^2=1$.  
Following the conventions in~\cite{Beneke:2019slt}, 
\begin{equation}
 \bar{S}^{(Q_{B})}_v(x) = \exp \left\{ + i e Q_{B} \int_{-\infty}^0 ds \, v \cdot A_{\rm us}(x + s v) \right\}   
\end{equation}
denotes the time-like Wilson line for the incoming 
$\bar{B}$ meson with four-velocity $v^\mu$ and charge 
$Q_{B}$. 
Charge conservation implies $Q_B = Q_{M_1} + Q_{M_2}$ 
in~\eqref{eq:usoftfac}, required to ensure the gauge invariance 
of the Wilson line product. The notation is general: for 
neutral mesons the corresponding Wilson line is simply unity.

This factorization can be shown by matching SCET$_{\rm II}$ 
non-perturbatively at the scale $\Lambda_{\rm QCD}$ to an 
effective theory of point-like mesons, which is, however, not the 
focus of this work. Nevertheless, the scale dependence of the 
non-radiative amplitude must match the scale dependence of the 
perturbative ultrasoft function. The logarithmic dependence on 
the radiated energy $\Delta E$ can be resummed rigorously in the 
limit $\Delta E \to 0$. Matching corrections at the scale of 
order $\Lambda_{\rm QCD}$, however, cannot be determined with 
perturbative methods.

The soft-photon-inclusive decay width is then given by 
\begin{equation}
\label{eq:usoftfct_rate}
    \Gamma[\bar{B} \to M_1 M_2](\Delta E) = |\mathcal{A}(\bar{B} \to M_1 M_2)|^2 \, \mathbf{\mathcal{S}}_\otimes(\{v_i\}, \Delta E) \,.
\end{equation}
The ultrasoft function
\begin{align}
 \mathbf{\mathcal{S}}_{\otimes}(\{v_i\}, \Delta E) =  \sum_{X_s} | \langle X_s | (\bar{S}_v^{(Q_{B})} S^{\dagger (Q_{M_1})}_{v_1} S^{\dagger (Q_{M_2})}_{v_2})(0) | 0 \rangle |^2 \, \theta(\Delta E - E_{X_s}) 
 \end{align}
accounts for the emission of an arbitrary number of 
ultrasoft photons (and electron-positron pairs) 
from the charged mesons with total 
energy $E_{X_s} \leq \Delta E$. 
At $\mathcal{O}(\alpha_{\rm em})$, and expanded 
to leading power in $m_{M_i} \ll m_B$, we find 
\begin{align}
\label{eq:oneloopus}
    \mathbf{\mathcal{S}}^{(1)}_{(+,-)} &= 8 \left(\frac12 +  \frac{1}{2}\ln\frac{m_{M_1}^2}{m_B^2} \right) \ln\frac{\mu}{2\Delta E} - \left(2+\ln\frac{m_{M_1}^2}{m_B^2}\right) \ln\frac{m_{M_1}^2}{m_B^2} - \frac{2}{3}\pi^2 \nonumber\\
& \hspace*{0.4cm}+ \,\,\, (m_{M_1}\to m_{M_2}) \\
    \mathbf{\mathcal{S}}^{(1)}_{(-,0)} &=  8 \left(1 + \frac{1}{2} \ln\frac{m_{M_1}^2}{m_B^2}\right) \ln\frac{\mu}{2\Delta E} - \left(2+\ln\frac{m_{M_1}^2}{m_B^2}\right) \ln\frac{m_{M_1}^2}{m_B^2} +4- \frac{2}{3}\pi^2  \ ,
\end{align}
and similarly for $\mathbf{\mathcal{S}}^{(1)}_{(0,-)}$ with 
$m_{M_1} \to m_{M_2}$. Obviously, $\mathbf{\mathcal{S}}^{(1)}_{(0,0)}=0$.
The expression for $\mathbf{\mathcal{S}}_{(+,-)}^{(1)}$
 is also given, e.g., in~\cite{vonManteuffel:2014mva}.

Although in this paper we provided the anomalous dimensions of the SCET operators, we leave the resummation of structure-dependent QED logarithms between the scales $m_B$ and $\Lambda_{\rm QCD}$ for future work. Since the scale ratio $\Lambda_{\rm QCD}/m_b$ is not extremely small, we do not expect the resummation of 
$(\alem \ln^2 m_b/\Lambda_{\rm QCD})^n$ terms to be important, and 
the fixed-order $\mathcal{O}(\alpha_{\rm em})$ expression 
should provide a very good approximation.
An exception are the logarithms in the ratio of the radiation energy cut $\Delta E \ll m_{M_i}$ and $m_B$, which can modify the rate at the level of a few percent.
These logarithms are universal in the sense that they can be extracted from the ultrasoft EFT with point-like mesons, or alternatively~\cite{Baracchini:2005wp} from scalar QED for point-like scalar mesons. 
When factorizing ultrasoft effects, the logarithms of $\mu/\Delta E$, 
which appear in (\ref{eq:oneloopus}),  must be related to IR 
($\mu_{\rm IR}$) 
scale dependence of the IR subtracted non-radiative amplitude 
(alternatively, the IR singularities of the unsubtracted on-shell 
amplitude). This dependence is contained in the (anti-) collinear and 
soft matrix elements, which define the QED-generalized LCDAs 
and form factors. Renormalization-group evolution from the hard 
scale $\mu_b$ to the collinear scale $\mu_c$ gives the 
universal Sudakov factors 
\begin{align}
\label{eq:collSudakov}
 e^{S_{M_i}(\mu_b, \mu_c)} = \exp \left\{ - \frac{\alpha_{\rm em}}{2\pi} \,Q_{M_i}^{\,2} \ln^2 \frac{\mu_c}{\mu_b} \right\} 
\end{align}
and a remainder, which defines the split of the leading double logarithms into 
this and the structure-dependent piece~\cite{Beneke:2019slt}. 
This separation is useful, because as shown below the above 
factor converts the scale $\mu$ in the exponentiated 
version of (\ref{eq:oneloopus}) into $m_B$, while the 
structure-dependent logarithms, which can depend on 
the charges of the constituents of the mesons rather than 
the mesons themselves, turn out to be small, at least for 
the case of $B_q\to \mu^+\mu^-$ considered in~\cite{Beneke:2019slt}. 
Here and below we work in the double-logarithmic approximation, 
except for logarithms in $\Delta E$. For the latter we include the 
full dependence as given in~\eqref{eq:oneloopus}.
At the level of the decay rate, the $\mu_c$ dependence of the 
factorized virtual $B\to M_1 M_2$ amplitude 
 cancels after taking into account ultrasoft emissions below $\Delta E$. Indeed, combining~\eqref{eq:collSudakov} with the exponentiated ultrasoft function evaluated at $\mu = \mu_c$, we find
\begin{eqnarray}
\label{eq:VRneutralB}
&& \left| e^{S_{M_1}(\mu_b,\mu_c) + S_{M_2}(\mu_b,\mu_c)} \right|^2 
e^{S_{\otimes}^{(1)}}
= \exp \bigg\{ \frac{\alpha_{\rm em}}{\pi}  
\,\bigg(Q_{B}^2+Q_{M_1}^2 
\bigg[1+\ln \frac{m_{M_1}^2}{m_B^2}\bigg]
\nonumber\\
&&\hspace*{1.5cm}  + \,Q_{M_2}^2 
\bigg[1+\ln \frac{m_{M_2}^2}{m_B^2}\bigg] \bigg) 
\,\ln \frac{m_B}{2 \Delta E}
 \bigg\} \,.
\end{eqnarray}
to the above mentioned accuracy. 
These results allow us to write the soft-photon-inclusive width  
with the large logarithmic dependence on the energy cut  
$\Delta E$ resummed to all orders in the standard form
\begin{align}
\label{eq:dEexponentiation}
    \Gamma[\bar B\to M_1 M_2](\Delta E) &= 
\Gamma^{(0)}[\bar B\to M_1 M_2]\ U(M_1 M_2) \,,
\end{align}
where 
\begin{equation}
\label{eq:Uusoft}
U(M_1 M_2) =  
\left(\frac{2\Delta E}{m_B}\right)^{-\frac{\alpha_{\rm em}}{\pi} 
\left(Q_{B}^2+Q_{M_1}^2 \Big[1+\ln\frac{m_{M_1}^2}{m_B^2}\Big]+
Q_{M_2}^2\Big[1+\ln\frac{m_{M_2}^2}{m_B^2}\Big] \right)}\,.
\end{equation}
Here $\Gamma^{(0)}$ is the square of the factorized virtual 
$B\to M_1 M_2$ amplitude discussed in earlier sections of 
this paper, with the universal Sudakov factors 
(\ref{eq:collSudakov}) divided out. 
There is an ambiguity in what one calls the ``non-radiative'' 
amplitude or decay width, but it is {\em this}~expression that 
most naturally deserves this name, given the universality 
and factorization-scale independence of 
the $\Delta E$ dependent radiation factors (\ref{eq:Uusoft}).
By definition, all large logarithms between 
the scale $m_B$ and $m_{M_i} \sim \Lambda_{\rm QCD}$ 
still contained in $\Gamma^{(0)}$ are structure-dependent logarithms 
whose resummation is not considered here. 

We close this section with a comparison of the treatment of 
soft-photon radiation in this section to the approach 
of~\cite{Baracchini:2005wp}. The authors express the 
soft-photon-inclusive decay width as the product of the 
non-radiative width and an energy-dependent correction 
factor $G_{12}(E)$, similar to~\eqref{eq:usoftfct_rate}. 
The precise definition of the non-radiative width is not 
specified, and $G_{12}(E)$ is computed from the virtual and 
real corrections in an effective theory that treats the 
$B$ meson and light mesons as point particles. 
Eq.~(5) in~\cite{Baracchini:2005wp} for $G_{12}(E)$ 
agrees with (\ref{eq:oneloopus}), if we put  $\mu=m_B$ 
in (\ref{eq:oneloopus}) and drop the virtual 
contributions $H_{12}$ and $N_{12}(\mu)$ to $G_{12}(E)$ 
in~\cite{Baracchini:2005wp}, as well as power-suppressed terms 
in $m_{M_i}/m_B$. Also, (\ref{eq:Uusoft})  is in agreement 
with~\cite{Baracchini:2005wp} in the appropriate limit 
$m_{M_i} \ll m_B$. 

There is nevertheless an important conceptual difference. 
Setting $\mu=m_B$ in (\ref{eq:oneloopus}) cannot 
be justified from the EFT of point-like mesons, since its 
UV scale of validity is at most $\Lambda_{\rm QCD}$. 
The treatment within SCET provided earlier in  
this paper is necessary to justify the neglect of structure-dependent 
logarithms such that one obtains (\ref{eq:Uusoft}) with the 
approximation (\ref{eq:collSudakov}). Conceptually, the main 
difference between the ultrasoft correction (\ref{eq:oneloopus}) 
and the function $G_{12}(E)$ in~\cite{Baracchini:2005wp} is, 
however, that the latter is defined in a theory with point-like 
light mesons, which are still {\em dynamical} degrees of 
freedom, whereas in our 
set-up, for photon energies much below $\Lambda_{\rm QCD}$, the 
light mesons are static and have only ultrasoft fluctuations,
similar to heavy quarks in heavy-quark effective theory. 
The logarithms of $m_{M_i}^2/m_B^2$ in the ultrasoft theory arise 
from the large boost of the rest frame of the light mesons 
relative to the $B$-meson rest frame. The ultrasoft function 
defined above receives no virtual correction in the one-loop 
approximation, because the integrals are scaleless, 
whereas the virtual corrections $H_{12}$ and 
$N_{12}(\mu)$ that enter $G_{12}(E)$ in the theory with 
dynamical point-like meson are non-zero, but not really 
meaningful. The reason is that keeping the mesons that have 
internal structure at distances of order $1/\Lambda_{\rm QCD}$ 
and masses of $\mathcal{O}(\Lambda_{\rm QCD})$ dynamical in 
a point-like description is inconsistent as the internal 
structure leads to higher-order multipole couplings that 
would give unsuppressed corrections to the virtual contributions 
when the internal loop momenta are of order $\Lambda_{\rm QCD}$, 
as is the case in~\cite{Baracchini:2005wp}. 
Fortunately, the virtual corrections are not needed to 
obtain the dependence of the ultrasoft radiation 
factors (\ref{eq:Uusoft}) on the resolution energy $\Delta E$, 
as was also recognized in~\cite{Baracchini:2005wp}, and 
hence the virtual correction there 
may be regarded as a contribution to the unspecified 
non-radiative amplitude in that framework.


\section{Estimates for $\pi K$ observables}
\label{sec:pheno}

Having set up the factorization, we present numerical estimates  of the QED effects. At this stage, we neither attempt an error analysis nor perform an analysis of all $B\to M_1 M_2$ decays but rather restrict ourselves to a first quantitative understanding of the QED effects for various $B\to \pi K$ decay observables that are often employed as diagnostics of New Physics. 
We distinguish three types of effects arising at different scales:
\begin{itemize}
    \item Electroweak scale to $m_B$: QED corrections to the Wilson coefficients
    \item $m_B$ to $\mu_c$: QED corrections to the hard-scattering kernels, form factors and decay constants
    \item below $\Lambda_{\rm QCD}$: Ultrasoft QED effects
\end{itemize}
The ultrasoft corrections only contribute at the level of the decay rate and will be discussed in more detail below. The QED corrections arising between the electroweak scale and $\mu_c$ can be interpreted as corrections to the colour-allowed tree-amplitude $\alpha_1(M_1M_2)$ and the colour-suppressed tree-amplitude $\alpha_2(M_1M_2)$, with an important caveat. In QCD, these amplitudes were introduced to factor out the hard-scattering kernels from the product of the universal form factors and decay constants defined by \cite{Beneke:2003zv}
\begin{align}
    A_{M_1 M_2}{}& \equiv i \frac{G_F}{\sqrt{2}} m_B^2 F_0^{BM_1}(0) f_{M_2} \ ,
    \end{align}
where $F_0^{BM_1}$ and $f_{M_2}$ are the standard QCD form factor and decay constant, respectively. Including QED effects requires the QCD$\times$QED generalized form of $A_{M_1M_2}$, which now depends on the charges $Q_{M_1}$ and $Q_{M_2}$:   
\begin{equation}\label{eq:Agen}
  \mathcal{A}(M_1M_2) \equiv  i \frac{G_F}{\sqrt{2}} m_B^2 \mathcal{F}_{Q_2}^{BM_1}(0) \mathscr{F}_{M_2} \ .
\end{equation}
In QED, both factors, $\mathcal{A}(M_1M_2)$ and $\alpha_{1,2}(M_1M_2)$, depend on the charges $Q_{M_1}$ and $Q_{M_2}$ and their separation is no longer compelling. Nevertheless, to stay as close as possible to the familiar notation, we can factor out the universal $A_{M_1M_2}$, and write 
\begin{equation}\label{eq:Aqcd}
\mathcal{A}(M_1M_2) \alpha_{i}(M_1M_2)  = 
A_{M_1M_2}\Big(\alpha_i^{\rm QCD}(M_1M_2) + \delta \alpha_{i}(M_1 M_2)\Big) \,,
\end{equation}
which puts all QED modifications into $ \delta \alpha_{i}(M_1 M_2)$. 
The $\mathcal{O}(\alem)$ QED correction $\delta \alpha_i$ is then a combination of different effects:
\begin{align}
   \delta \alpha_{i}(M_1M_2) &\equiv \delta \alpha_i^{\rm WC}(M_1M_2)+ \delta \alpha_i^{\rm K}(M_1M_2) + \delta \alpha_i^{\rm F, V}(M_1M_2) +\delta \alpha_i^{\rm F, sp}(M_1M_2) \,.
\end{align}
The four terms stem from QED corrections to the Wilson coefficients (WC), hard and hard-collinear scattering kernels (K), and form factors and decay constants of the vertex (F,V) and spectator (F,sp) terms, respectively. The latter two also contain the QED corrections to the LCDAs. When estimating the QED corrections numerically below, we restrict ourselves to $\mathcal{O}(\alem)$ only. Since spectator scattering first occurs at $\mathcal{O}(\alem, \alpha_s)$, $\delta\alpha^{\rm F,sp}$ is $\mathcal{O}(\alem \alpha_s)$ and will thus be dropped. We also neglect the vertex correction $\delta\alpha^{\rm F,V}$, 
since the QED effects on the form factors and decay constants are 
not (yet) known. For charged $M_2$ decays, the situation is a bit more involved. Recall that in that case, we replace $\mathcal{F}_{-}^{BM_1}$ with the semi-leptonic amplitude according to \eqref{eq:semilepFF}, which introduces the semi-leptonic Wilson coefficient $C_{\rm sl}$ and the leptonic factor $Z_\ell$, which contribute to $\delta \alpha^{\rm WC}$ and $\delta\alpha^{\rm F, V}$, respectively. Since we neglect the latter, we must also set $Z_\ell=1$. However, since we do include the QED effects in the Wilson coefficients, we have to account for $C_{\rm sl}$ in decays to charged $M_2$ mesons.  

\subsection{Electroweak corrections to the Wilson coefficients}

To obtain the QED correction to the Wilson coefficients, we follow \cite{Huber:2005ig} (see also \cite{Bobeth:2003at}), where QCD logarithms are summed but QED logarithms are not. Including the summation of QED logarithms would be technically more challenging while their effect is small. A consequence of not summing the QED logarithms is that we obtain an expansion in  $\alpha_s$ and $\kappa \equiv \alem/\alpha_s$. The expansion in $\kappa$ instead of $\alem$ itself arises from the fact that all powers of $c_s=\alpha_s L$, where $L$ is a large logarithm, are summed. Explicitly, this entails that all logarithmically enhanced QED terms $\alem  L = c_s \alem/\alpha_s$ get replaced by $f(c_s)\alem/\alpha_s$, where $f(c_s)$ is found by solving the renormalization-group equation (RGE) \cite{Huber:2005ig}. Therefore, $\kappa$ is the natural expansion parameter. 

Since we are interested in the leading QED corrections, we only consider corrections of $\mathcal{O}(\kappa, \kappa \alpha_s)$. We note however that also the $\mathcal{O}(\kappa \alpha_s^2)$ terms are available \cite{Huber:2005ig}. The QED correction to the Wilson coefficients $C_1$ and $C_2$ are obtained at NLL, that is, by including the $\alem$ corrections to the $C_i(\mu_0)$, the $\mathcal{O}(\kappa, \kappa \alpha_s)$ corrections to the anomalous dimension, and the three- and four-loop pure QCD contributions to the running of $\alpha_s$ and $\alem$. The RGE is then solved perturbatively in terms of
\begin{equation}
    \lambda \equiv \frac{\beta_{0}^{\rm em} \alem(\mu_0)}{\beta_{0}^s \alpha_s(\mu_0)} \,,  \quad\quad\quad \omega \equiv 2 \beta_{0}^s \,
\frac{\alpha_s(\mu_0)}{4\pi} \,,
\end{equation}
where $\beta_{0}^s=\tfrac{23}{3}$ and $\beta_{0}^{\rm em}=\tfrac{80}{9}$ at $n_f=5$. The Wilson coefficients can then be written as
\begin{equation}
    C_i(\nu) = C_i^{\rm QCD}(\nu) + \delta C_i(\nu) \ ,
\end{equation}
where the $\delta C_i$ contain the $\alem$ corrections. The pure QCD NNLL Wilson coefficients and the coupling constants $\alpha_s$ and $\alem$, for which we use the $\overline{\rm{MS}}$ scheme and initial conditions at $m_Z$, are listed in Table~\ref{tab:inputs}. The couplings for $n_f=5$ at $\nu=4.8$ GeV are also specified there. We find
\begin{eqnarray}
\delta C_1(\nu) &=& -1.66\frac{\alem(\nu)}{4\pi} =
-1.00\cdot 10^{-3}\,,\\[0.5mm]
\delta C_2(\nu) &=&5.68 \frac{\alem(\nu)}{4\pi} = 3.42 \cdot 10^{-3}\ .
\end{eqnarray}
We can now compute the QED effect on the tree amplitude coefficients 
from the Wilson coefficients, which gives 
\begin{align}
    \delta\alpha_1^{\rm WC}(M_1 M_2) &= \delta C_2  =5.68\frac{\alem(\nu)}{4\pi} = 3.42 \cdot 10^{-3}\ ,\\
       \delta\alpha_2^{\rm WC}(M_1 M_2) &= \frac{4}{9} \delta C_1 + \frac{1}{3} \delta C_2 = 1.16 \frac{\alem(\nu)}{4\pi}  =0.695 \cdot 10^{-3} \ .
\end{align}
There is still one subtle point. For charged $M_2$, the replacement of the form factor by the semi-leptonic amplitude introduces the Wilson coefficient $C_{\rm sl}$. Its one-loop fixed-order expression is \cite{Sirlin:1981ie}
\begin{equation}\label{eq:csl}
   \delta C_{\rm sl}(\nu) = \frac{\alem(\nu)}{\pi} \ln \frac{m_Z}{\nu} =  11.78 \frac{\alem(\nu)}{4\pi} = 7.09 \cdot 10^{-3} \,.
\end{equation}
As we will show, for $B\to \pi K$ decays, the charged $M_2$ decays only have contributions from $\alpha_1$. Therefore, in fact, we must use
 \begin{equation}
     \delta \alpha_1^{\rm WC}(M_1 M_2) = \delta C_2 - \delta C_{\rm sl}\; C_2^{\rm tree}=  -3.88 \cdot 10^{-3}\,,
 \end{equation}
where for consistency we neglect $\mathcal{O}(\alpha_s \alem)$ terms and use $C_2^{\rm tree}(\nu)=1.03$. Interestingly, the normalization to the semi-leptonic amplitude changes the sign of $\delta\alpha^{\rm WC}_1$, but its magnitude remains similar.

\subsection{QED contributions from the hard-scattering kernels}
The QED contribution to the colour-allowed and colour-suppressed coefficients $\alpha_1$, $\alpha_2$ are
\begin{equation}
   \delta\alpha_{i}^{\rm K}(M_1M_2)=\frac{\alem(\mu)}{4\pi} \sum_{j=1,2} 
C_j^{\rm QCD}(\nu) \left[  \mathscr{V}_j^{(1)}(M_2) + H_{j, Q_{2}}^{\rm em}(M_1M_2) \right].
\end{equation}
The convolution of the LCDA of $M_2$ with the hard-scattering kernel 
$T^{\rm {I}}_{i,Q_2}$ is defined by
\begin{equation}
  \mathscr{V}_i(M_2)= \int_0^1 du\; T_{i,Q_2}^{\rm{I}}(u)
\,\phi_{M_2}(u) \,.
\end{equation}
For $\phi_{M_2}$ we use the standard Gegenbauer expansion,  
recalling that we neglect all QED corrections to non-perturbative 
objects such as the LCDA and approximate them by their QCD values. For neutral $M_2$, keeping only the first two Gegenbauer coefficients, we find 
\begin{align}
    \mathscr{V}^{(1)}_2(M_2^0) & =
    -\frac{2}{27}\left[- 6 L_\nu -18  -3 i\pi + \left(\frac{11}{2}-3i\pi\right)a_1^{M_2} - \frac{21}{20}a_2^{M_2} \right]
    \end{align}
and $\mathscr{V}^{(1)}_{1}(M_2^0) = C_F \mathscr{V}^{(1)}_{2}(M_2^0)$. While for charged $M_2$, we find 
\begin{align}\label{eq:V2}
    \mathscr{V}^{(1)}_{2}(M_2^-)  = &
\left[-\frac{5}{3}L-\frac{2L_\nu}{3}-\frac{97}{18}-\frac{22 i \pi }{9}-\frac{\pi ^2}{9} \right. \nonumber \\
 & \left. -\left(\frac{1}{2}L+\frac{133}{72}+\frac{i \pi}{3} \right) a_1^{M_2}  - \left(\frac{3 }{5}L+\frac{184}{75}+\frac{2 i \pi }{5}\right)  a_2^{M_2} \right]\ ,
\end{align}
and   $\mathscr{V}^{(1)}_{1}(M_2^-)=0$. We note that we reduced the QED correction by introducing the semi-leptonic amplitude, which cancelled some of the double logarithms present in the hard-scattering kernels $H^{\rm I}_{i,-}(u)$ in \eqref{eq:H2miphys}. 
As discussed previously, the $\nu$ dependence from $L_\nu$ gets cancelled by the Wilson coefficients (including  $C_{\rm sl}$), while the $\mu$ dependence cancels against the QED scale dependence of $\mathscr{F}_{M_2} \Phi_{M_2}/Z_\ell$. However, as we do not take QED corrections to this quantity into account in our numerical estimates, the $\mu$ dependence from $\mathscr{V}_2^{(1)}(M_2)$ remains. 
For the spectator-scattering terms, we obtain 
\begin{align}
 H_{2,-}^{\rm em}(M_1 M_2)  &=
\frac{4\pi^2 Q_{sp} Q_u}{N_c} \frac{r_{\rm sp}(M_1)}{9} \int_0^1 du\; dv\; \frac{\phi_{M_2}(u) \phi_{M_1}(v)}{\bar{u}\bar{v}} \ , \\
 &=\frac{4\pi^2 Q_{sp} Q_u r_{\rm sp}(M_1)}{N_c}\sum_{i,j} a_i^{M_1} a_j^{M_2}  \ ,
\end{align}
where $a_j^{M_i}$ is the $j$th Gegenbauer moment for the meson $M_i$ 
(with $a_0^{M_i}\equiv 1$ in QCD) and
\begin{equation}
   r_{\rm sp}(M_1) \equiv \frac{9 f_B f_{M_1}}{m_B \lambda_B 
F_0^{BM_1}(0) } \ .
\end{equation}
The other charge combinations are related to $H_{2,-}^{\rm em}(M_1 M_2)$, similar to the relations between the $T_{i,(Q_1,Q_2)}^{\rm II}$ in \eqref{eq:Trelations}, by
\begin{equation}
 H_{1,-}^{\rm em}(M_1 M_2)= 0\,, \quad 
H_{1,0}^{\rm em}(M_1 M_2)=C_F H_{2,0}^{\rm em}(M_1 M_2) = \frac{C_F}{N_c}H_{2,-}^{\rm em} (M_1 M_2)\,.
\end{equation}

We note that the Wilson coefficients are evaluated at the scale $\nu$, while $\mathscr{V}_i^{(1)}(M_2)$ depends on both scales $\nu$ and $\mu$. As we sum QCD, but not QED logarithms, the question arises what scale should be taken for $\alem$ and for the QCD parameters (i.e.~the Gegenbauer coefficients of the light mesons and $\lambda_B$). In principle, several choices could be justified. In the following analysis, we take $\mu=1\,$GeV at the collinear scale. To obtain $\alem(\mu=1 \GeV)$, we use the one-loop RG evolution, include the quark flavour thresholds at $4.8 \GeV \,(n_f=4), 1.2 \GeV \,(n_f=3)$, and the decoupling of the $\tau$ lepton at $\mu_\tau=1.78\,$GeV. Values are given in Table~\ref{tab:inputs}. 
\begin{table}[t]
	\begin{center}
		\begin{tabularx}{0.98\textwidth}{|c C C C|}
	\hline 
	\multicolumn{4}{|c|}{Coupling constants and masses [GeV]}\\
		\hline 
  $\alem(m_Z)=1/127.96$  & $\alpha_s(m_Z) = 0.118$ & $m_B = 5.297$ & $m_Z=91.19$ 
  \\ \hline
		\end{tabularx}
				\vskip 1pt
\begin{tabularx}{0.98\textwidth}{|C C C C C|}
\hline
	\multicolumn{5}{|c|}{Decay constants [MeV] and form factors}\\
		\hline 
$f_\pi= 130$  & $f_K=160$  & $f_B=190$  & $F_0^{B\pi}=0.25$ & $F_0^{BK}=0.34$  \\  	\hline
\end{tabularx}
\vskip 1pt
\begin{tabularx}{0.98\textwidth}{|C C|}
\hline
\multicolumn{2}{|c|}{CKM parameters and $R_{\pi K}$}  \\ \hline
$|\lambda_u/\lambda_c| \equiv  |V_{us} V_{ub}^* /V_{cb}V_{cs}^*|= 0.0206$  &  $R_{\pi K} = f_\pi F_0^{BK}/f_K F_0^{B\pi}=1.11$\\ 
\hline
\end{tabularx}
\vskip 1pt
	\begin{tabularx}{0.98\textwidth}{|C C C C|}
	\hline
	\multicolumn{4}{|c|}{Wilson coefficients and coupling constants at $\nu=4.8$ GeV}\\
		\hline 
  $C_1^{\rm QCD}=-0.26$ & $C_2^{\rm QCD}=1.01$    & $\alem=1/132.24$  & $\alpha_s = 0.216  $ 
  \\ \hline
		\end{tabularx}
		\vskip 1pt
		\begin{tabularx}{0.98\textwidth}{|C C C C|}
\hline	\multicolumn{4}{|c|}{Parameters of distributions amplitudes at $\mu=1$ GeV}\\
		\hline 	
$a_2^{\pi}=0.138$ & $a_1^{\bar K}=0.061$ & $a_2^{\bar{K}} = 0.124$ & $\lambda_B = 250$ MeV \\ \hline
\end{tabularx}
		\vskip 1pt
		\begin{tabularx}{0.98\textwidth}{|C C |}
\hline	\multicolumn{2}{|c|}{Coupling constants and $\hat{\alpha}_4^c$ at $\mu=1$ GeV}\\
		\hline 	
 $\alem=1/134.05$  & $\hat{\alpha}_4^c = -0.104- 0.015 i$ \\ \hline
\end{tabularx}
\caption{Inputs for the estimate of the QED effects. The Gegenbauer coefficients are taken from \cite{Bali:2019dqc} and evolved to $1\,$GeV with  LL accuracy. The pure QCD Wilson coefficients are evaluted at 
the NNLL order.}
\label{tab:inputs}
	\end{center}
\end{table}

\subsubsection{Penguin-dominated $B\to \pi K$ decays}

The $B\to \pi K$ decay amplitudes are given by \cite{Beneke:2003zv}
\begin{align}
    \mathcal{A}_{B^-\to \pi^- \bar{K}^0} & = A_{\pi K} \hat\alpha_4^p \nonumber \ , \\
       \sqrt{2} \mathcal{A}_{B^-\to \pi^0 K^-} & = A_{\pi K} \left[\delta_{pu} \alpha_1 +\hat\alpha_4^p\right] + A_{K\pi} \left[\delta_{pu} \alpha_2 + \delta_{pc} \frac{3}{2} \alpha_{3, \rm EW}^c\right] \nonumber \ , \\
           \mathcal{A}_{\bar{B}^0\to \pi^+ K^-} & = A_{\pi K} \left[\delta_{pu} \alpha_1 + \hat\alpha_4^p \right]\nonumber \ , \\
       \sqrt{2} \mathcal{A}_{\bar{B}^0\to \pi^0 \bar{K}^0} & = A_{\pi K} \left[-\hat\alpha_4^p\right] + A_{K\pi} \left[\delta_{pu} \alpha_2 + \delta_{pc} \frac{3}{2} \alpha_{3, \rm EW}^c\right] \nonumber \ , 
\end{align}
where the $\alpha_i$ carry the argument $(M_1M_2)$. Here, $\hat{\alpha}_4$ and $\alpha_{3, \rm EW}^c$ are QCD (electroweak) penguin coefficients as defined in \cite{Beneke:2003zv}. In addition, each term is multiplied with the CKM factor $V_{pb}V^*_{ps}$ and summed over $p=u,c$. 
Due to the unique association of the right and wrong insertion 
with the charge factors and $\alpha_{1,2}$, we find 
\begin{align}
\delta \alpha_{1}^{\rm K}(\pi^+ K^-)& = \frac{\alem(\mu)}{4\pi} C_2^{\rm QCD} \left[\mathscr{V}_2(K^-)+ H_{2,-}^{\rm em}(\pi^+ K^-)\right] , \\
\delta \alpha_{1}^{\rm K}(\pi^0 K^-)& = \delta \alpha_{1}^{\rm K}(\pi^+ K^-) + \frac{\alem(\mu)}{4\pi}\Delta_1^{\rm K} , \\
\delta \alpha_{2}^{\rm K}(\bar{K}^0 \pi^0)&= \frac{\alem(\mu)}{4\pi} (C_F C_1^{\rm QCD} + C_2^{\rm QCD}) \left[\mathscr{V}_2(\pi^0) + H_{2,0}^{\rm em}(\bar{K}^0\pi^0)\right] , \\
\delta \alpha_{2}^{\rm K}(K^- \pi^0)&= \delta \alpha_{2}^{\rm K}(\bar{K}^0 \pi^0) +\frac{\alem(\mu)}{4\pi} \Delta_2^{\rm K}\ .
\end{align}
Since the vertex corrections $\mathscr{V}_i$ do not depend on the charge of $M_1$, only spectator scattering contributes to the difference between the two charge configurations of $\delta\alpha_{1,2}$, 
defined by
\begin{align}
    \Delta_1^{\rm K} & = C_2^{\rm QCD}(\nu)\; \Big( H_{2,-}^{\rm em}(\pi^0K^-)-H_{2,-}^{\rm em}(\pi^+ K^-)\Big) = 8.03\; \frac{r_{\rm{sp}}(\pi)}{0.674}\ ,\\
\Delta_2^{\rm K} & = \left(C_F C_1^{\rm QCD}(\nu)+C_2^{\rm QCD}(\nu)\right)  \Big(H_{2,0}^{\rm em}(K^-\pi^0)- H_{2,0}^{\rm em}(\bar{K}^0\pi^0)\Big)
=1.59\;\frac{r_{\rm{sp}}(K)}{0.610}  \,.
\end{align}
Finally, the hard-scattering kernel contributions 
to $\delta\alpha_{i}^{\rm K}$ are
\begin{align}
 \delta\alpha_{1}^{\rm K}(\pi^+ K^-)& = \frac{\alem(\mu)}{4\pi}\left[-0.89-7.96 i - 2.68 \;\frac{r_{\rm{sp}}(\pi)}{0.674}\right] =(-2.12- 4.73  i)\cdot 10^{-3}\,, \\
  \delta\alpha_{1}^{\rm K}(\pi^0 K^-)& = \frac{\alem(\mu)}{4\pi}\left[-0.89 - 7.96 i + 5.36 \;\frac{r_{\rm{sp}}(\pi)}{0.674}\right] =(2.65 - 4.73  i)\cdot 10^{-3}\,, \\
  \delta\alpha_{2}^{\rm K}(\bar{K}^0 \pi^0)&=\frac{\alem(\mu)}{4\pi}\left[0.83 + 0.46 i -0.53  \;\frac{r_{\rm{sp}}(K)}{0.610}\right] = (0.18 + 0.27 i)\cdot 10^{-3} \,,\\
 \delta\alpha_{2}^{\rm K}(K^- \pi^0)&=\frac{\alem(\mu)}{4\pi}\left[0.83 + 0.46 i +1.06 \;\frac{r_{\rm{sp}}(K)}{0.610}\right] = (1.12 + 0.27 i)\cdot 10^{-3} \,.
\end{align}
The numerical values are at the per mille level. 
As discussed previously, there is a logarithmic 
$\mu$ dependence in $\delta\alpha_i$, which should be cancelled by that of $\mathscr{F}_{M_2} \Phi_{M_2}/Z_\ell$, but is not in our approximation of neglecting QED effects on the hadronic quantities. 
Changing the collinear scale to $\mu=1.5$ GeV, changes the 
real part of the form-factor term (first number in the square bracket) by $\mathcal{O}(1)$.
We will show below, however, that this ambiguity drops out when considering ratios of branching fractions or direct CP asymmetries.


\subsection{Ultrasoft factors}
\label{sec:pheno:us}

When considering branching ratios also ultrasoft effects should be taken into account. This is done simply by multiplying the rate with $U(M_1M_2)$ defined in \eqref{eq:Uusoft}. Here $\Delta E$ is the window of the $\pi K$ invariant mass around $m_B$. For our theory to be valid, we require $\Delta E \ll \Lambda_{\rm QCD}$. Similar as in the $B_q\to \mu^+\mu^-$ analysis \cite{Beneke:2019slt}, we adopt $\Delta E = 60\,$MeV, which defines the signal window. In recent experimental analyses, such a signal window is not used (only a cut on the invariant mass of $5\,$GeV is employed) and the mass spectrum is modelled using PHOTOS to account for the photon radiation (see e.g. \cite{Carbone:2010wza, Aaij:2018tfw}). In order to compare theory with experiment, it is beneficial to perform the experimental analysis within a signal window as above, such that no extrapolations are necessary. 
Numerically, the ultrasoft factors are relatively important: 
\begin{align}
U(\pi^+ K^-) & = 0.914 \ , \nonumber \\
U(\pi^0 K^-)&=  U(K^-\pi^0)  = 0.976 \ , \nonumber \\
U(\pi^- \bar{K}^0) &= 0.954 \ , \nonumber \\
U(\bar{K}^0 \pi^0) & = 1 \ . 
\end{align}
For decays to charged $M_2$, the situation is again more involved due to the replacement of the generalized form factor by the non-radiative semi-leptonic amplitude, such that
\begin{equation}
{\rm Br}(\pi^+K^-) \propto \big|\,\mathcal{A}_{\rm non-rad}^{\rm sl, M_1} \,\alpha_1(\pi^+K^-)\big|^2 \,U(\pi^+ K^-) \ , \end{equation}   
and similar for ${\rm Br}(\pi^0 K^-)$. The non-radiative semi-leptonic rate is itself obtained from the 
branching ratio
\begin{equation}\label{eq:usoftcorsl}
{\rm Br}(M_1 \ell^-) =  U(M_1 \ell^-) \,|\mathcal{A}_{\rm non-rad}^{\rm sl, M_1} |^2\ ,
\end{equation}
where the ultrasoft function differs from $U(M_1 K^-)$ only due to the mass difference between $\ell^-$ and $K^-$. In the following, we assume that the ultrasoft correction in \eqref{eq:usoftcorsl} was applied to the semi-leptonic rate such that $\mathcal{A}_{\rm non-rad}^{\rm sl, M_1}$ was determined 
and employed in the calculation of the non-radiative non-leptonic amplitude.

\subsection{Ratios, isospin 
sum rule, and CP asymmetries}

Adding the three sources of QED effects discussed above, 
gives sub-percent corrections to the branching fractions from the hard-scattering kernels and Wilson coefficients, 
and potentially larger ultrasoft radiation effects for final states 
with charged particles. Therefore, it is more interesting to study ratios of decay rates in which 
QCD corrections are suppressed. To this extent, we first consider
\begin{align}
    R_L &= \frac{2 \,{\rm Br}(\pi^0 \bar{K}^0)+ 2\,{\rm Br}(\pi^0 K^-)  }{ {\rm Br}(\pi^- \bar{K}^0) + {\rm Br}(\pi^+ K^-) }  = R_L^{\rm{QCD}} + \delta R_L \ .
    \end{align}
The QCD part is given by
\begin{equation}
         R_L^{\rm{QCD}} =    1 +   |r_{\rm EW}|^2 - \cos\gamma \;{\rm Re}\;(r_{\rm T} r_{\rm{EW}}^*) + \ldots \ , \nonumber \\
    \end{equation}
    where $r_{\rm EW} \,(r_{\rm T})$ are ratios of electroweak penguin coefficients $\alpha_{3,\rm EW}$ (tree coefficient $\alpha_1$) over the dominant QCD penguin coefficient $\hat{\alpha}_4^c$, which are typically $\mathcal{O}(0.1)$ \cite{Beneke:2003zv}. Therefore, $R_L^{\rm QCD}$ was expanded in these small ratios, and the dots represent higher-order or negligible terms in this expansion. We observe that the QCD corrections to unity enter only quadratically in these small ratios. QED effects, however, enter 
linearly:
    \begin{equation}
  \delta R_L     = \cos\gamma \;{\rm Re}\,(\delta_E)  + \delta_{\rm U} \ .
    \end{equation}

The QED correction $\delta_E$  comes from the hard-scattering kernels and the Wilson coefficients. We already mentioned that only the spectator-scattering contribution depends on the charge of the $M_1$ meson. Therefore, in the ratio $R_L$ only the difference between the spectator-scattering terms, denoted by $\Delta_{i}^{\rm K}$, contributes at leading order. For this reason,  $\delta_E$ is $\mu$ independent 
at $\mathcal{O}(\alem)$ and does not suffer from the uncancelled 
$\mu$ dependence discussed previously. In fact, as the correction to the Wilson coefficients does not depend on the charge of $M_1$ either, it also does not contribute at this order, and we find 
\begin{equation}\label{eq:delE}
    \delta_{E} = \frac{\alpha_{\rm em}(\mu)}{4\pi}\left|\frac{\lambda_u}{\lambda_c}\right|\frac{\Delta^{\rm K}_1+ \Delta^{\rm K}_2 R_{\pi K}}{\hat{\alpha}_4^c(\pi K)} = \left(-1.89 + 0.27 i\right)\frac{\alpha_{\rm em}(\mu)}{4\pi}=(-1.12 + 0.16  i)\,\cdot 10^{-3} \ ,
\end{equation}
where we used the CKM ratio $\lambda_u/\lambda_c$, form-factor ratio $R_{\pi K}$ and $\hat{\alpha}_4^c$ given in Table~\ref{tab:inputs}. 
The contribution from the hard-scattering kernels to $\delta R_L$ 
is seen to be 
at the per mille level, and gets suppressed by the cosine of 
the CKM angle $\gamma$. The ultrasoft factors give the 
$\mathcal{O}(r^0)$ correction
\begin{equation}
    \delta_U \equiv \frac{  1  + U(\pi^0 K^-)}{U(\pi^- \bar{K}^0)  + U(\pi^+ K^-) } -1 = 
5.8 \% 
\end{equation}
in the expansion in small amplitude ratios. Contrary to the kernel 
correction, the ultrasoft correction depends on $\Delta E$, which 
in turn depends on how the measurement is performed.  
Finally, combining both terms and using $\gamma=70^\circ$, we find
\begin{equation}
      \delta R_L   = 5.7 \% \,, 
\end{equation}
which is dominated by the ultrasoft effect. This should 
be compared to the smaller QCD correction \cite{Beneke:2003zv} 
$R_L^{\rm QCD} -1 = 0.01\pm 0.02$.

Besides ratios of branching fractions, also CP asymmetries form interesting observables. Using isospin relations, a sum rule
\begin{align}
    \Delta(\pi K)& \equiv A_{\rm CP}(\pi^+K^-) + \frac{\Gamma(\pi^- \bar{K}^0)}{\Gamma(\pi^+K^-)} A_{\rm CP}(\pi^- \bar{K}^0) - \frac{2\Gamma(\pi^0 K^-)}{\Gamma(\pi^+K^-)} A_{\rm CP}(\pi^0 K^-) \nonumber \\
    & - \frac{2\Gamma(\pi^0 \bar{K}^0)}{\Gamma(\pi^+K^-)} A_{\rm CP}(\pi^0 \bar{K}^0) \equiv  \Delta(\pi K)^{\rm {QCD}}  + \delta\Delta(\pi K) 
\end{align}
between the CP asymmetries of the different $\pi K$ decays was 
identified that should exhibit only small deviations from 
zero \cite{Gronau:2006xu, Gronau:2005kz}. Indeed, the 
pure QCD part is
\begin{align}
     \Delta(\pi K)^{\rm {QCD}}     & = 2 \sin\gamma\, \left[{\rm Im}\left(r_T r_{\rm EW}^*\right) +2 \,{\rm Im}\left(r_C r_{\rm EW}^*\right) 
\right] + \ldots  ,
\end{align}
where we have again expanded in the small amplitude ratios and the dots represent higher-order or negligible terms. The phase of $\alpha_{3,\rm{EW}}^c$ approximately equals that of $\alpha_1$, such that the first term is suppressed. Therefore, the QCD contribution is dominated by the interference between the colour-suppressed tree amplitude $r_{C}$ and the electroweak penguin contribution $r_{\rm EW}$, resulting in $\Delta(\pi K)^{\rm {QCD}}= (0.5\pm 1.1)\%$ \cite{Bell:2015koa}. The QED correction enters linearly and can be 
expressed in terms of the ultrasoft contribution $\delta\Delta_U$ and the same $\delta_E$ from \eqref{eq:delE}, 
but this time only the imaginary part enters:
\begin{equation}
    \delta\Delta(\pi K)=   -2 \sin\gamma\, {\rm Im}\,(\delta_E)  
+ \delta\Delta_U \ .
\end{equation}
As the $\Delta_i^{\rm K}$  are real (we only consider tree-level spectator scattering at $\mathcal{O}(\alem)$), the imaginary part of $\hat{\alpha}_4^c$ drives this contribution, which turns out to be negligible. There is no $\mathcal{O}(r^0)$ ultrasoft contribution to the sum rule, since at this order all CP asymmetries vanish. The first 
non-vanishing term in the expansion in small amplitude ratios is 
\begin{align}
\delta\Delta_U = & \,2 \sin\gamma\, \Big[\, {\rm Im}\,(r_P- r_T)+{\rm Im}\,(r_P)\frac{U(\pi^-\bar{K}^0)}{U(\pi^+ K^-)} \nonumber \\
& + {\rm Im}\,(r_T+r_C-r_P) \frac{U(\pi^0 K^-)}{U(\pi^+ K^-)}   - \frac{{\rm Im}\,(r_P+ r_C)}{U(\pi^+ K^-)} \,\Big] = -0.39 \% ,
\end{align}
where we used $r_C = 0.06 - 0.016i$, $r_P= 0.018 + 0.0038i$ and 
$r_T= 0.18 - 0.030i$ defined as in  \cite{Beneke:2003zv}. 
This factor is sensitive to the imaginary parts of QCD parameters, which are difficult to determine with high precision, hence $\delta\Delta_U$ may suffer from a relatively large uncertainty. The combined QED effect is
\begin{equation}
    \delta\Delta(\pi K)=  -0.42 \% \ ,
\end{equation}
which is similar in size to the QCD correction, so that the isospin 
CP asymmetry sum rule is not only robust against QCD contributions, 
but also free from sizeable QED contaminations. 
To conclude this discussion, we also give the QED corrections to
the individual CP asymmetries. In first order in the small 
amplitude ratios, the QED effect is a linear shift 
$\delta A_{\rm CP}$ of the QCD-only result. The ultrasoft 
factors always cancel in individual CP asymmetries as they 
are the same for the decay rate and its CP conjugate.
We then find  
\begin{align}
\delta A_{\rm CP}(\pi^+K^-) & = 
2 \sin\gamma   \left|\frac{\lambda_u}{\lambda_c}\right|
\,{\rm Im} \,\frac{\delta \alpha_1(\pi^+ K^-)}{\hat\alpha_4^c(\pi K)}
= 0.14\%\,, 
\nonumber \\
\delta  A_{\rm CP}(\pi^- \bar{K}^0) & = 0\,,  
\\
\delta A_{\rm CP}(\pi^0 \bar{K}^0) & = 
-2 \sin\gamma   \left|\frac{\lambda_u}{\lambda_c}\right| 
R_{\pi K}\,{\rm Im}\,
\frac{\delta \alpha_2(\bar{K}^0\pi^0 )}{\hat\alpha_4^c(\pi K)}
=0.01\%\,, 
\nonumber \\
\delta  A_{\rm CP}(\pi^0 K^-) & = 2 \sin\gamma  \left(\, 
\left|\frac{\lambda_u}{\lambda_c}\right|\,{\rm Im}\left[
\frac{\delta \alpha_1(\pi^+ K^-) + R_{\pi K}\delta \alpha_2(\bar{K}^0\pi^0 )}{\hat\alpha_4^c(\pi K)}\right] +\,{\rm Im}\, 
\delta_E \right) =0.16\%\,,
\nonumber \end{align}
where $\delta\alpha_{1,2}$ now contain both $\delta\alpha^{\rm K}$ and $\delta\alpha^{\rm WC}$. 
Finally, we can consider the difference
\begin{equation}
    \delta(\pi K) \equiv A_{\rm CP}(\pi^0 K^-) - A_{\rm CP}(\pi^+ K^-)
\end{equation}
 between the two CP asymmetries with a charged final-state kaon, 
which receives the tiny QED correction 
\begin{equation}
2 \sin\gamma  \left( \left|\frac{\lambda_u}{\lambda_c}\right| R_{\pi K} {\rm Im}\;\frac{\delta \alpha_2( \bar{K}^0\pi^0)}{\hat\alpha_4^c(\pi K)} +{\rm Im}\;\delta_E \right) = 0.02 \% \,.
\end{equation}
All of these QED corrections are much smaller than the QCD 
uncertainties.

\section{Conclusion}
\label{sec:conclusion}

The question whether QCD factorization of non-leptonic charmless 
two-body decays can be extended to include QED effects has been 
investigated here for the first time. Any attempt to include 
QED effects mandates the precise definition of an observable 
that includes soft photon radiation, since in general the 
final-state mesons can be electrically charged. We considered 
the soft-inclusive decay rates $\Gamma[\bar{B} 
\to M_1 M_2 + X_s]\big\vert_{E_{X_s} \leq \Delta E}$,
where the final state $X_s$ consists of photons and possibly also 
electron-positron pairs with total energy less than 
$\Delta E \ll \Lambda_{\rm QCD}$ in the $B$-meson rest frame.
Factorization then refers to purely virtual 
electromagnetic effects on scales from $m_B$ to a few times 
$\Lambda_{\rm QCD}$. Electromagnetic effects above $m_B$ 
can be conceptually trivially included in the Wilson 
coefficients of the effective weak interactions, those 
below a few times $\Lambda_{\rm QCD}$ in hadronic matrix elements, 
suitably generalized for QED effects.

Our first main result consists in the statement that the 
non-leptonic two-body decay amplitudes can indeed be factorized in 
a way such that the QCD factorization formula \eqref{eq:QCDF} 
retains its original form, but the hard and hard-collinear 
scattering kernels now receive QCD and QED corrections, 
which can be computed in perturbation theory. Despite this 
similarity in form, the physics contained in the short-distance 
kernels is nevertheless more involved than in QCD alone, 
since the second meson $M_2$ does not decouple completely from 
the $B\to M_1$ transition. When $M_2$ is 
electrically charged, soft virtual photon exchange 
leads to a dependence of the generalized hadronic matrix 
elements on light-like Wilson lines that ``remember'' the 
directions of flight and charges of the particles. To our knowledge, 
we provide the first definition of light-meson LCDAs 
including QED effects. The interpretation of these is subtle.
The generalized $B$-meson LCDA in turn should rather be considered 
as the soft function for the process, which by its definition 
contains the soft rescattering physics of the process. 
Calculating these hadronic matrix elements with non-perturbative 
methods appears challenging 
for the time being, but at least the precise definitions of the 
required matrix elements can now be given.

Second, we computed the QED short-distance coefficients 
at leading order in the electromagnetic coupling. Their 
IR finiteness checks the validity of the factorization formula 
at this order. We then provided first quantitative 
estimates of QED corrections to the $\pi K$ final states, 
for which QCD-insensitive ratios of branching fractions 
and CP asymmetry sum rules are prime targets for 
precision measurements in high-luminosity $B$ physics 
experiments. In these estimates we include on top of  
the QED corrections from the kernels, which were the focus 
of this work, the effect from the Wilson coefficients 
and ultrasoft radiation. The latter depend on the experimental 
set-up and might reach a few percent, but the former two 
were found to be at the sub-percent to per mille level. 
To a certain extent this is fortunate, since, as noted 
above, a consistent treatment of all QED effects should 
also include the presently unknown effects in the 
generalized hadronic matrix elements. 

We point out that there remains a gap in our understanding 
of QED effects at the hadronic scale, which is related 
to the interpretation of the QED-generalized decay constants, 
form factors and LCDAs, which are all ``non-radiative'' objects. 
As defined here they are technically IR divergent---their 
IR divergences cancel with the IR divergences in ultrasoft 
real emission, which can be computed in a theory of 
point-like hadrons. A proper interpretation of the QED-generalized 
decay constants, form factors and LCDAs can be given as 
matching coefficients to the ultrasoft theory, where 
fluctuations at the $\Lambda_{\rm QCD}$ scale have been 
integrated out. However, this matching will have to be 
defined and computed non-perturbatively. Similar problems 
are presently addressed in lattice QCD/QED for electromagnetic 
corrections to leptonic and semi-leptonic decays of light mesons 
\cite{Giusti:2017dwk,Sachrajda:2019uhh,deDivitiis:2019uzm, Kane:2019jtj}. Nevertheless, the present problem appears to be a formidable 
challenge for lattice calculations, as the operators to be 
computed involve light-like Wilson lines.

\subsubsection*{Acknowledgements}
We thank Christoph Bobeth, Tobias Huber, Stefano Perazzini and 
Robert Szafron for discussions. This research was supported by the 
DFG Sonderforschungsbereich/Transregio 110 ``Symmetries and the Emergence of Structure in QCD''. J.-N. T. would like to thank 
the ``Studienstiftung des deutschen Volkes'' for a scholarship.


\appendix

\section{\boldmath Photon polarization and $\bar{B}_q^0\to M_1^+ M_2^-$ spectator scattering} 
\label{app:sec:spec}

\begin{figure}
\centering
\includegraphics[scale=0.95]{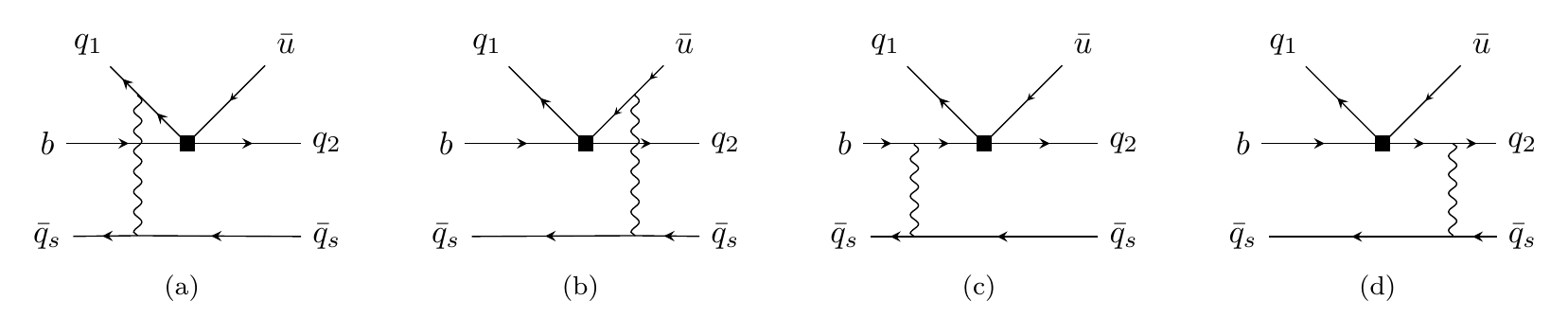}
\includegraphics[scale=0.95]{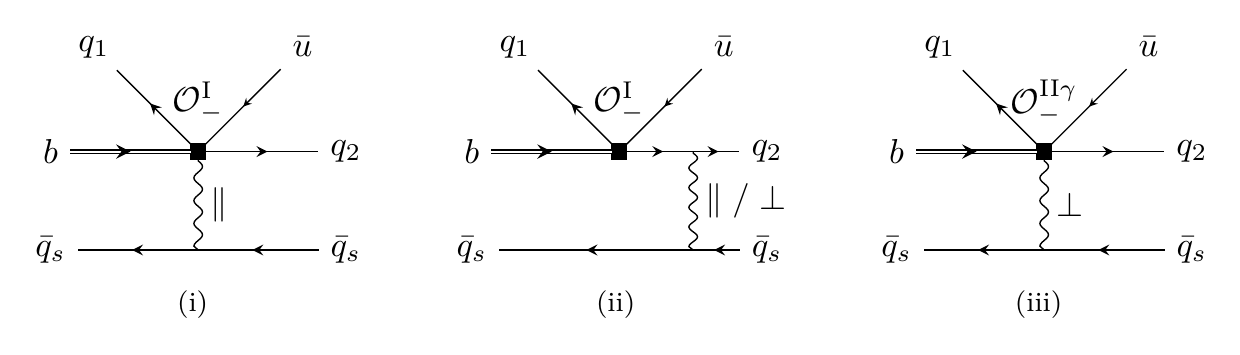}
\caption{Tree-level spectator-scattering in the full theory and in 
SCET$_{\rm I}$.}
\label{fig:spec_scatt_old} 
\end{figure}

In Section~\ref{subsec:H2} we stated that the tree-level scattering 
kernels $H_{i,-}^{\rm II \gamma}$ are fully determined by the first 
three diagrams in Fig.~\ref{fig:diaII} with a transversely 
polarized external photon. Here we provide more details on this 
important fact, as it guarantees that the spectator-scattering term 
in the factorization formula is free from endpoint divergences 
even when the meson $M_2$ is electrically charged. In particular, 
we show that the 
spectator scattering through longitudinally polarized photons as 
well as the full contribution from the last diagram in 
Fig.~\ref{fig:diaII}, which would both be endpoint divergent, are 
exactly recovered by certain time-ordered products of the 
operator $\mathcal{O}^{\rm I}_-$. Hence both are correctly included 
in the non-perturbative QED-generalized form factors.

For this purpose, it is instructive to compute the 
spectator-scattering diagrams (a) -- (d) shown  in 
Fig.~\ref{fig:spec_scatt_old} in the full theory,
as well as the SCET$_{\rm I}$ diagrams (i) -- (iii),  
with the LCDA projector method as in the original 
QCD factorization works~\cite{Beneke:1999br,Beneke:2000ry}.
In this method, we relate the partonic amplitudes to hadronic matrix 
elements defining the heavy- and light-meson LCDAs by replacing 
the on-shell spinors with certain projectors. For the 
case of spectator scattering, this amounts to integrating out hard and 
hard-collinear modes simultaneously, matching directly to 
SCET$_{\rm II}$. Including QCD contributions up to twist-3, 
but applying the so-called Wandzura-Wilczek approximation, 
which neglects three-particle LCDAs at twist-3, 
the projector for the $B$-meson operating on a partonic 
amplitude with spinors stripped off and spinor (colour) 
indices $\beta\alpha$ ($ba$) 
is given by~\cite{Beneke:2000wa}
\begin{eqnarray}
\label{bproj}
M^B_{\alpha\beta} &=& 
- \frac{i f_B m_B}{4}\,\frac{\delta_{ab}}{N_c}
\,\Bigg[ \frac{1+\slashed{v}}{2} \Bigg\{
\phi^B_+(\omega)\,\slashed{n}_+ +  \phi^B_-(\omega)\,\slashed{n}_- 
\nonumber\\
&& \hspace*{2cm}
- \,\int_0^{\omega} d\eta \,\left(\phi^B_-(\eta)-
\phi^B_+(\eta)\right)\,\gamma^\mu\frac{\partial}{\partial 
l_{\perp}^\mu}\Bigg\} \, \gamma_5 \Bigg]_{\alpha\beta}.
\end{eqnarray}
Following the notation of \cite{Beneke:2003zv}, we have for 
light pseudoscalar mesons 
\begin{eqnarray}\label{pimeson2}
   M_{\alpha\beta}^P &=& \frac{i f_P}{4} \,\frac{\delta_{ab}}{N_c}
\,\Bigg[
   \slashed{p}\,\gamma_5\,\phi_P(x) 
\nonumber\\ 
&& \hspace*{0cm}
- \,\mu_M\gamma_5 \left(
   \phi_p(x) - i\sigma_{\mu\nu}\,\frac{p^\mu\,\bar p^\nu}{p\cdot\bar p}\,
   \frac{\phi'_\sigma(x)}{6}
   + i\sigma_{\mu\nu}\,p^\mu\,\frac{\phi_\sigma(x)}{6}\,
   \frac{\partial}{\partial k_{\perp\nu}} \right)\!\Bigg]_{\alpha\beta}\!\!,
\end{eqnarray}
where $p$ is the momentum of the meson, $\bar{p}$ is a light-like 
vector, whose three-components point in the opposite direction 
of $p$, and the transverse derivatives act on the 
quark momenta in the partonic amplitude. For the present purposes 
it is sufficient to identify the LCDAs with those in QCD alone, 
but the projector method would also work for the QED-generalized 
LCDAs. Although we work to leading power, it is instructive 
to keep the twist-3 two-particle LCDAs for the following 
reason. The subleading twist-3 LCDAs $\phi^B_-$, $\phi_p$ and 
$\phi_\sigma$ enter the heavy-to-light form factors at leading 
power~\cite{Beneke:2000wa} with endpoint-divergent convolutions, 
which is the reason why the matrix element of the SCET$_{\rm I}$ 
operator $\mathcal{O}^{\rm I}$ is not matched to SCET$_{\rm II}$. 
It is therefore important to understand the twist-3 terms as well 
for the QED spectator-scattering effects in the 
non-leptonic factorization formula. 

To disentangle the polarization components 
of the internal photon line 
in Fig.~\ref{fig:spec_scatt_old}, we decompose the metric tensor 
into its longitudinal and transverse parts, $g^{\mu\nu}= (n_{+}^{\mu}n_{-}^{\nu} + n_{-}^{\mu} n_{+}^{\nu})/2 + g_\perp^{\mu\nu}$.
The leading-power full-theory results for the individual diagrams 
(a) -- (d) and polarization state in Feynman gauge are 
\begin{align}
\begin{aligned}
\langle Q_2 \rangle^{a)}_\parallel &= - Q_d Q_{\rm sp}  \left\langle {\bar{v}^{-2}} \right\rangle_{M_1} \left\langle \omega^{-1} \right\rangle_-  \, , \quad &\langle Q_2 \rangle^{a)}_\perp = &~0 \, , \\
\langle Q_2 \rangle^{b)}_\parallel &=  Q_u Q_{\rm sp}  \left\langle{\bar{v}^{-2}} \right\rangle_{M_1} \left\langle{\omega}^{-1} \right\rangle_-  \, , &\langle Q_2 \rangle^{b)}_\perp = &  \,  Q_u Q_{\rm sp}  \left\langle {\bar{v}^{-1}} \right\rangle_{M_1} \left\langle {\bar{u}^{-1}} \right\rangle_{M_2}  \left\langle {\omega}^{-1}\right\rangle_+ \, , \\
\langle Q_2 \rangle^{c)}_\parallel&=  Q_d Q_{\rm sp} \left\langle {\bar{v}^{-2}} \right\rangle_{M_1} \left\langle{\omega}^{-1} \right\rangle_-  \, , &\langle Q_2 \rangle^{c)}_\perp =&  \,  Q_d Q_{\rm sp} \left\langle {\bar{v}}^{-1} \right\rangle_{M_1} \left\langle {\omega}^{-1} \right\rangle_+\, , \nonumber \\[-0.3cm]
\end{aligned}
\end{align}
\begin{eqnarray}
\langle Q_2 \rangle^{d)}_\parallel &=& 
Q_u Q_{\rm sp}  \frac{\mu_{M_1}}{3} \left\langle {\bar{v}^{-2}} \right\rangle_{\sigma 1} \left\langle {\omega^{-2}} \right\rangle_+ \ ,
\nonumber\\
\langle Q_2 \rangle^{d)}_\perp &=&  Q_u Q_{\rm sp} \left\langle {\bar{v}}^{-1}\right\rangle_{M_1}\left\langle {\omega}^{-1} \right\rangle_- + Q_u Q_{\rm sp}  \frac{\mu_{M_1}}{3} \left\langle {v^{-1}\bar{v}^{-1}} \right\rangle_{\sigma 1} \left\langle {\omega^{-2}} \right\rangle_+ \ ,
\label{HSproj}\end{eqnarray}
where we set $Q_{q_1}=Q_d, Q_{q_2}=Q_u$ for $\bar{B}_q^0\to M_1^+ M_2^-$ decays, and $Q_{\rm sp}$ is the charge of the spectator quark $q_s$. We factored out the overall normalization
\mbox{$\mathcal{N}\equiv i\pi\alpha f_{M_1} f_{M_2} f_B m_B/N_c$}, 
and defined
\begin{align}\begin{aligned}
   \left\langle v^n \right\rangle_{X} &\equiv \int_0^1 dv\, v^n \phi_{X}(v) \, , \quad  &\left\langle \omega^n \right\rangle_\pm &\equiv \int_{0}^\infty d\omega\, \omega^n \phi^B_\pm(\omega) \, .
   \end{aligned}
\end{align}
The sum of all terms constitutes the matrix element of the 
left-hand side of the matching relation~\eqref{eq:matchI}.

The endpoint behaviour of the various LCDAs implies that 
$ \left\langle {\bar{v}}^{-2}\right\rangle_{M},  \left\langle {\bar{v}}^{-2}\right\rangle_{\sigma}$, $\left\langle {\omega^{-2}} \right\rangle_+$, $\left\langle {\omega^{-1}} \right\rangle_-$ are ill-defined 
(divergent). Hence we observe that diagrams (a)~--~(c) result in 
divergent convolutions but only if the exchanged photon is 
longitudinally  polarized, while in diagram (d) also the 
transverse photon polarization leads to ill-defined convolutions.
In the QCD-alone treatment of spectator scattering, the corresponding gluon exchanges in diagrams (c) and (d) are absorbed into the $B \to M_1$ transition form factor and never considered explicitly, whereas the gluon attachments (a), (b) to the emitted meson $M_2$ sum up to 
zero. For photon exchange the situation is 
different. Diagrams (a) and (b) sum up to a divergent contribution that is proportional to the total charge of the $M_2$ meson from which one might conclude that for charged $M_2$ the second term in the factorization theorem is ill-defined, leading to a breakdown of factorization.
Fortunately, as already discussed in the main text, this is not the case since the longitudinal photon contributions arise from the hard-collinear Wilson line in the operator $\mathcal{O}^{\rm{I}}_-$ and are thus also associated with the ``form-factor term", 
which is never matched to SCET$_{\rm II}$.

To demonstrate this explicitly, we compute the SCET$_{\rm I}$ matrix elements of $\mathcal{O}^{\rm I}_-$ (diagrams (i) and (ii) in Fig.~\ref{fig:spec_scatt_old}) and $\mathcal{O}^{\rm II \gamma}_-$ (diagram (iii) in Fig.~\ref{fig:spec_scatt_old}) on the right-hand side of the matching relation~\eqref{eq:matchI}, 
projecting onto the same meson LCDAs as the full-theory diagrams.
We obtain for matrix elements of the  momentum-space 
operators
\begin{eqnarray}
 \langle \mathcal{\wt{O}}^{\rm{I}}_-(u) \rangle
&\equiv&  \int \frac{d\hat{t}}{2\pi} \, e^{-i u \hat{t}} \, 
\langle \mathcal{O}^{\rm{I}}_-(t) \rangle
\nonumber \\
&& \hspace*{-1cm} 
= \,\mathcal{N} Q_u Q_{\rm sp} \, \phi_{M_2}(u) \Big[ \left\langle {\omega}^{-1} \right\rangle_-  \left\langle {\bar{v}^{-2}} + {\bar{v}^{-1}} \right\rangle_{M_1}
    + \frac{\mu_{M_1}}{3} \left\langle {v^{-1}\bar{v}^{-2}} \right\rangle_{\sigma 1} \left\langle{\omega^{-2}} \right\rangle_+ \! \Big] \,,
    \end{eqnarray}
    and
\begin{eqnarray}
 \langle \mathcal{\wt{O}}^{\rm{II}\gamma}_-(u,v) \rangle 
\equiv 
  \int \frac{d\hat{s}}{2\pi} \, \frac{d \hat{t}}{2\pi} \, e^{-i (u \hat{t} + (1-v) \hat{s})} \, \langle \mathcal{O}^{\rm{II}\gamma}_-(t,s) \rangle 
  =  \mathcal{N} \,\frac{Q_{\rm sp}}{2} \, \frac{\phi_{M_1}(v)}{\bar{v}} \, \phi_{M_2}(u) \left\langle {\omega}^{-1} \right\rangle_+ \, .
\quad
\end{eqnarray}
Comparing to the full-theory result~\eqref{HSproj}, and given that $H^{\rm{I}}_{2,-}(u) = 1 + \mathcal{O}(\alpha_s,\alpha_{\rm em})$ has already been determined from the matching of the 
$\mathcal{O}^{\rm I}_-$ operator in four-quark matrix elements, 
we indeed find that all endpoint-divergent moments are contained in the matrix element of $\mathcal{O}^{\rm I}_-$, i.e. in the generalized soft form factor $\zeta^{B M_1}_{Q_{2}}$.
Further, we can read off the matching coefficient
\begin{align}
H^{\rm{II}\gamma}_{2,-}(u,v) = \frac{2Q_u}{\bar{u}} + 2Q_d \, ,
\end{align}
in agreement with~\eqref{eq:HII} from the direct matching of the 
operator with a transverse photon field only.
For completeness, we give the relations between the full-theory diagrams and individual SCET diagrams:
\begin{eqnarray}
   \langle Q_2 \rangle^{a)+b)+c)}_\parallel &=& \int_0^1 du~H^{\rm{I}}_{2,-}(u) \langle \mathcal{\wt{O}}^{\rm{I}}_-(u) \rangle^{i)} \, , 
\nonumber\\
 \langle Q_2 \rangle^{d)}_\parallel + \langle Q_2 \rangle^{d)}_\perp  &=& \int_0^1 du~ H^{\rm{I}}_{2,-}(u) \langle \mathcal{\wt{O}}^{\rm{I}}_-(u) \rangle^{ii)}  \, , \nonumber\\ 
   \langle Q_2 \rangle^{a)+b)+c)}_\perp  &=& \int_0^1 dvdu~H^{\rm{II}\gamma}_{2,-}(u,v)  \langle \mathcal{\wt{O}}^{\rm{II}\gamma}_-(u,v) \rangle^{iii)}  \, . \label{diagrel2}
\end{eqnarray} 
These results show once more that only transverse photons from the first three QED diagrams contribute to $H^{\rm{II} \gamma}_{2,-}(u,v)$.


\providecommand{\href}[2]{#2}\begingroup\raggedright\endgroup

\end{document}